\documentclass[twocolumn,showpacs,preprintnumbers,amsmath,amssymb]{revtex4}

\usepackage{graphicx}
\usepackage{dcolumn}
\usepackage{bm}

\begin{document}

\preprint{APS/123-QED}

\title{A local field emission study of partially aligned carbon-nanotubes
by AFM probe}

\author{A. Di Bartolomeo}
 \email{dibant@sa.infn.it}
\author{A. Scarfato}%
\author{F. Giubileo}%
\author{F. Bobba}%
\author{M. Biasucci}%
\author{A.M. Cucolo}%
\affiliation{%
Dipartimento di Fisica "E.R: Caianiello", Universit\`a degli Studi
di Salerno, and CNR-INFM Laboratorio Regionale SUPERMAT, and INFN
Gruppo collegato di Salerno, via S. Allende 84081 Baronissi (SA),
Italia}%

\author{S. Santucci}
\author{M. Passacantando}
 \affiliation{
Dipartimento di Fisica, Universit\`a dell'Aquila and INFN, via
Vetoio 67010 Coppito (AQ), Italia
}%

\date{\today}

\begin{abstract}
We report on the application of Atomic Force Microscopy (AFM) for
studying the Field Emission (FE) properties of a dense array of
long and vertically quasi-aligned multi-walled carbon nanotubes
grown by catalytic Chemical Vapor Deposition on a silicon
substrate. The use of nanometric probes enables local field
emission measurements allowing investigation of effects non
detectable with a conventional parallel plate setup, where the
emission current is averaged on a large sample area. The
micrometric inter-electrode distance let achieve high electric
fields with a modest voltage source. Those features allowed us to
characterize field emission for macroscopic electric fields up to
250 V/$\mu$m and attain current densities larger than 10$^5$
A/cm$^2$. FE behaviour is analyzed in the framework of the
Fowler-Nordheim theory. A field enhancement factor $\gamma
\approx$ 40-50 and a turn-on field $E_{turn-on} \sim$15 V/$\mu$m
at an inter-electrode distance of 1 $\mu$m are estimated. Current
saturation observed at high voltages in the I-V characteristics is
explained in terms of a series resistance of the order of
M$\Omega$. Additional effects as electrical conditioning, CNT
degradation, response to laser irradiation and time stability are
investigated and discussed.

\end{abstract}

\pacs{68.37.-d; 73.63.Fg; 81.07.De}
\maketitle

\section{Introduction}

Controlled propagation of electrons in vacuum is at the basis of
several  technological applications, like CRT displays, vacuum
electronics, electron microscopy, X-ray generation, electron beam
lithography, etc. The most common technique to extract electron
from matter is thermionic emission, which requires a source heated
at high temperature ($\sim 1000 ^{\circ}$C) and has several
drawbacks. Field emission (FE), which involves extraction of
electrons from a conducting solid (metal or highly doped
semiconductor) by an external electric field, is becoming one of
the best alternatives. Indeed, by this method, an extremely high
current density with low energy spread of the emitted electrons
and with negligible power consumption can be achieved
\cite{coldcathode1,cc2,cc3}.

High macroscopic electric fields of several kV/$\mu$m are required
for electrons to tunnel through the surface-to-vacuum potential
barrier. Such high fields can be practically obtained by
exploiting the local electric field enhancement at the apex of a
tip with small radius of curvature. With pointed cathodes, the
macroscopic electric field needed for electron emission can be
reduced to few V/$\mu$m.

For their high aspect ratio (diameter in the nanometer scale and
length of several microns), extremely small radius of curvature,
unique electric properties, high chemical stability and mechanical
strength, carbon nanotubes (CNTs) \cite{CNT1,CNT2,CNT3,CNT4,CNT5}
can be extraordinary field emitters and interest in their
applicability for FE devices has been steadily growing since their
discovery in 1991 \cite{discovery}.

Field emission has been observed both from single-walled (SWCNT)
and multi-walled (MWCNT) carbon nanotubes, individual or in an
ensemble (CNT films). Current densities over 1 A·cm$^{-2}$ at
macroscopic applied fields of few V/$\mu$m have been reported
\cite{fiveyears1,fy2,fy3,fy4}.

In technological applications, films of vertically aligned
nanotubes grown or deposited on a substrate have the largest
potential: the fabrication is relatively easy and suitable for
industrial production, the patterning is possible through optical
or electron beam lithography, and, depending on the inter-tube
spacing \cite{spacing}, extremely high emission capability can be
attained \cite{vertical1,v2,v3,v4,v5}.

Over the past ten years, a  variety of vertical CNT films,
differing for tube type (MW or SW), shape, dimension, density,
substrate, etc. have been extensively studied and used in
prototype FE devices, like displays, lamps, X-ray tubes, microwave
power amplifiers, etc. \cite{Application1,A2,A3,A4,A5}. However,
the unavoidable inhomogeneous composition and morphology of CNT
films, even on micrometric areas, make the comprehension of the
influence of fabrication and structural parameters on the FE
properties a scientific challenge.

Experiments with phosphor
screens have evidenced that emission from a CNT film originates
from isolated spots. FE current measurements performed by
conventional large area anode setups are affected by the dominant
contribution of a small subset of highly emitting CNTs that can
hide important characteristics of the remaining majority
\cite{subset1,ss2}. Therefore, small (possibly nanometric) area anodes are
essential for an accurate investigation of individual
emission site current-voltage (I-V) or current-time (I-t)
characteristics and for obtaining statistical data on the
spatially or time dependent FE behaviour of thin film emitters.

In this report we present a detailed study of FE performed on a
film of partially aligned MWCNTs with an anode consisting of a
nanometric AFM/STM probe in a high vacuum chamber. A large amount
of experimental data, on different sites of a single sample,
allowed a significant statistic analysis of several effects. In
the majority of the measurements, we observed a reproducible FE
current saturation at high fields, which is explained in terms of
a series resistance modified Fowler-Nordheim model. Emission
stability and response to radiation, relevant topics for
technological applications, are discussed.

\section{CNT production}

A dense array of vertical and partially aligned carbon nanotubes
was produced by catalytic Chemical Vapor Deposition (CVD). A
Nickel film of 3 nm was deposited on a silicon substrate (001,
p-type, $\rho =1-40 \Omega$cm), covered by a thin layer of SiO$_2$
(thickness $\sim$3 nm), acting as diffusion layer and preventing
the formation of NiSi$_x$ (which does not have a catalytic
function). Growth of carbon nanotubes was achieved by chemical
decomposition of acetylene on clusterized Ni (catalyst) upon
heating at about 700 $^{\circ}$C in NH$_3$ ambient. Acetylene was
injected in the chamber together with ammonia in the ratio
C$_{2}$H$_{2}$/NH$_{3}$ 1:5. The gas ratio, the chamber
temperature and the growth time were the main control parameters.

The average height of the CNTs produced with this procedure was
around 15 $\mu$m as can be seen from the SEM images shown in Fig.
1. Vertical alignment is due to a crowding effect, i.e.
neighbouring tubes supporting each other by van der Waals forces.
A TEM analysis (not shown here) revealed that the nanotubes are
multi-walled with inner and outer diameters  of 5-10 nm and 15-25
nm, respectively.

\begin{figure}
\includegraphics[width=7.5cm]{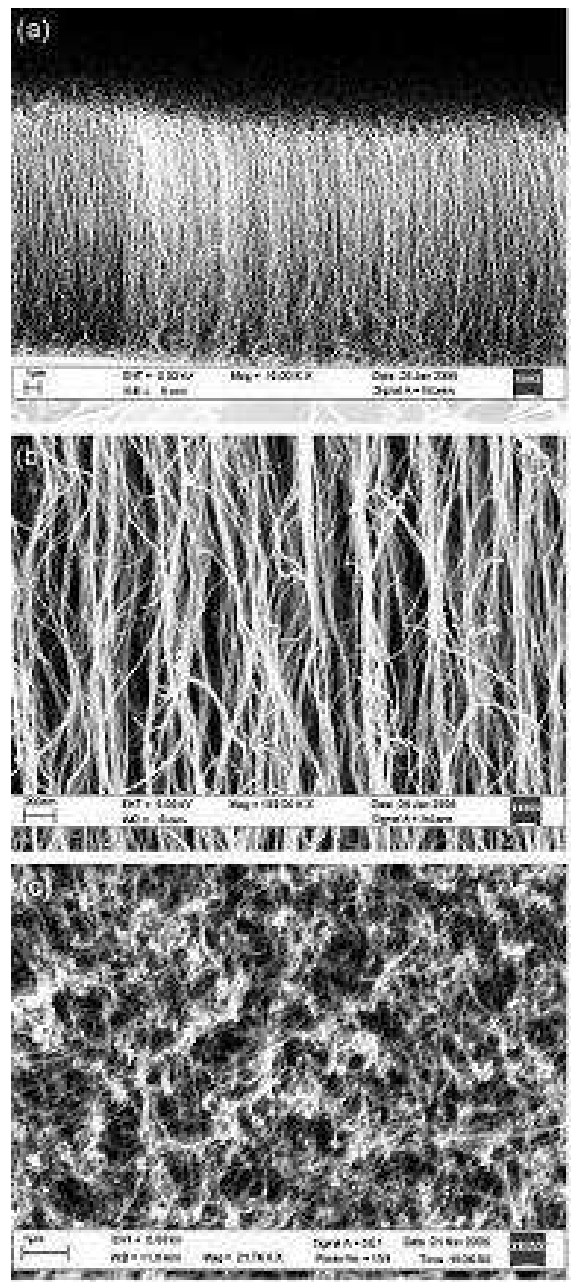}
\caption{\label{fig:epsart} SEM images of vertically aligned
MWCNTs, grown by catalytic chemical vapour deposition on a silicon
substrate (area 5$\times$5 mm$^2$)  with Ni catalyst. (a), (b)
lateral views, (c) top view.}
\end{figure}

SEM top view (Fig. 1c) shows also that our film lose verticality
in the upper part towards the air surface. This is a non-critical
limitation for field emission applications since the nanotubes are
stretched and aligned along the cathode-anode direction at
typically a quarter of the voltage necessary to trigger emission,
regardless of their initial disposition
\cite{saturation1,saturation2}.

Particles, that backscattered SEM analysis proved to have a high
Ni content, are visible at the upper end of the nanotubes,
indicating a weak catalyst adherence to the surface and a
dominating "tip growth" mechanism \cite{growth}.

\section{Measurement  setup}

Field Emission measurements were performed by means of an Omicron
UHV STM/AFM system, operating at room temperature, and connected
to a Semiconductor Parameter Analyser Keithley 4200-SCS, working
as SMU (source-measurement unit). A detailed scheme for FE
apparatus is shown in Fig. 2.

Current measurements were carried out initially in a static vacuum
of $10^{-3} - 10^{-4}$ mbar produced by a turbo-molecular pump,
and successively, to improve the signal to noise ratio, at
$10^{-7} - 10^{-8}$ mbar by an ionic pump installation. High
vacuum is crucial to reduce the effects of adsorbates as O$_{2}$,
H$_{2}$, H$_{2}$O, N$_{2}$ on FE current value and stability (it
was shown for example that H$_{2}$O increases the emission while
O$_{2}$ decreases it dramatically \cite{H2O}; any gas enhances
fluctuations).

\begin{figure}
\includegraphics[width=8.0cm]{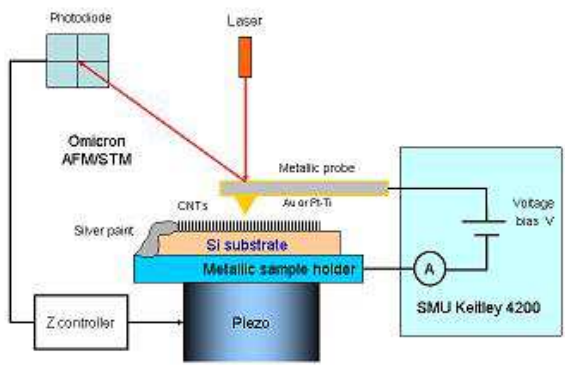}
\caption{\label{fig:epsart} AFM probe used as counter-electrode
for the measurement of the FE current from a vertically aligned
array of MWCNTs. The whole apparatus consists of a vacuum chamber
hosting an AFM/STM connected to an external source measurement
unit.}
\end{figure}

\begin{figure}
\includegraphics[width=5.0cm]{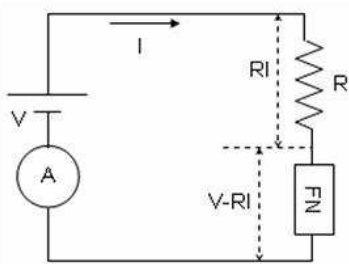}
\caption{\label{fig:epsart} Schematic representation of the
measurement circuit. A is the ammeter, V the voltage source, R the
series resistance and FN is the Fowler-Nordheim current device,
constituted by the CNT film emitter array and the AFM tip.}
\end{figure}

\begin{figure}
\includegraphics[width=6.5cm]{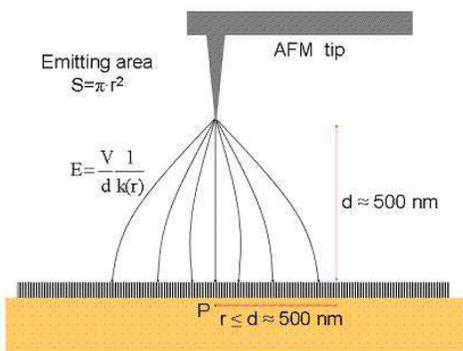}
\caption{\label{fig:epsart} : Electric field in the region between
an AFM tip and a flat CNT emitting surface.}
\end{figure}

A two probe method was chosen for its simplicity. The electrical
connection of the CNT film and the Si substrate with the metallic
sample holder was assured  through a spot of silver paint.The
equivalent circuit is modelled in Fig. 3, where R is a series
resistance accounting for CNTs, interfaces, contacts and wires
resistances. Voltage sweeps were performed on the allowed range of
-210 to +210 V. The current flowing through the tip was measured
with an accuracy better that 1 pA.

Horizontal movements of the tip, with steps of $\sim$0.1 nm,
covered an area of $5\times 5 \mu$m$^2$ and accurate vertical
control was over a distance of 2 $\mu$m, with resolution better
than 0.1 $\AA$.The AFM tips (Au- or Pt/Ti-coated polysilicon), of
conical shape, had aperture of $\sim 30^{\circ}$, curvature radius
$\leq$ 35 nm, height 20-25 $\mu$m. The elastic constant of the
cantilever was $\sim$40 N/m and the resonant frequency $\sim$300
kHz.

In the literature, FE investigations are mostly carried out in a
parallel plate setup, with diode or triode configuration, where
current is averaged on a large sample area. Our setup, exploiting
a very pointed anode, enables FE measurements over a limited
circular region, whose radius is shorter than the tip-film
distance $d$ (see following). By taking into account the average
CNT diameter and density, a maximum of 50 to 250 CNTs, for  $d$ =
500 nm, are expected to take part in the emission process.

In addition, the (sub)micrometric values of $d$, let a
(macroscopic) field of a few hundreds V/$\mu$m be attained with a
low voltage source and electrons can be extracted by fields
significantly higher than those of few tens V/$\mu$m commonly
used.

\section{Field Emission Theory and Simulation}

\subsection{Fowler-Nordheim theory}

FE occurs when electrons from a solid tunnel through the surface
potential barrier whose width is reduced by the application of an
external electric field. The emission current depends on the
electric field at the emitter surface (referred as microscopic or
local electric field), $E_S$, and on the workfunction, $\Phi$,
i.e. the effective surface-vacuum barrier height. The
Fowler-Nordheim model \cite{FNtheory}, derived for a flat
metallic surface at 0 $^{\circ}$K and assuming a triangular
potential barrier, predicts an exponential behaviour of the
emitted current:

\begin{equation}
  I=S \cdot a\frac{E_S^2}{\Phi}exp\left(-b\frac{\Phi^{3/2}}{E_S}\right)
\end{equation}

where $S$ is the emitting surface area, E$_S$ is the uniform
electric field on that surface and $a$ and $b$ are constants. When
$S$ is expressed in cm$^2$, $\Phi$ and E$_S$ respectively in eV
and V/cm$^2$, $a = 1.54\cdot 10^{-6}$ AV$^{-2}$eV and $b =
6.83\cdot 10^{7}$ eV$^{-3/2}$V cm$^{-1}$.

In a parallel plate configuration, the field E$_S$ can be obtained
from the applied potential V and the inter-electrode distance $d$
as $E=V/d$. If the cathode is constituted by an array of sharp
tips, a field enhancement factor, $\gamma$, which takes into
account the amplification occurring around their apexes, has to be
introduced and

\begin{equation}
  E_S= \gamma \frac{V}{d}
\end{equation}

According to (1) and (2), a Fowler-Nordheim plot of ln(I/V$^2$) as
a function of 1/V  is a straight line, whose slope, $m=b\Phi
^{3/2}d/ \gamma$, and interception, $y_0=ln[aS\gamma^2 / (\Phi
d^2)]$, can, in principle, be used to estimate $\gamma$ and $\Phi$
\cite{phenomenological}.

Although corrections \cite{corrections1,corrections2} are required
to describe effects of non zero temperature, series resistance,
extremely curved surfaces and non-uniform field enhancement
factors or workfunctions, the basic FN theory has proven to
be a good model to achieve a first-approximation understanding of
the emission phenomena. For temperatures up to several hundred
degree Celsius and fields in a large window, F-N model provides a
good fitting to the I-V characteristics of several kind of
emitters, included individual or in-film carbon nanotubes.

Due to our setup geometry, E$_S$ is non-uniform, depending on the
distance $r$ from the point just below the tip apex (point P in
Fig. 4). We will show that we can overcome this complication by
introducing an effective emitting area $S$ and tip correction
factor $k$, whose values will be determined by means of a
numerical simulation.

\subsection{Electric field and FE behaviour}

The electric field generated by an ideal conical tip, of the
dimensions of our AFM probe at a bias voltage V, on a grounded
flat graphite surface at distance d, was numerically calculated by
MAXWELL \cite{MAX}, a software which solves electromagnetic
problems by finite element analysis. Graphite was chosen for its
conductivity comparable to that of a CNT film.

The result obtained for $d$ = 500 nm and V = 150V is shown in Fig.
5, where we can observe the electric field lines (top) and
magnitude levels (bottom).
\begin{figure}
\includegraphics[width=7.5cm]{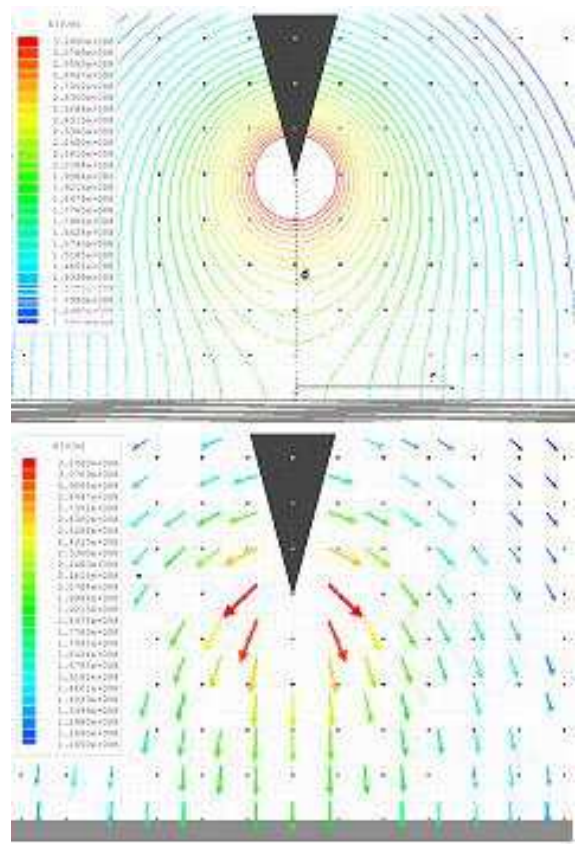}
\caption{\label{fig:epsart} Electric field (vector and magnitude
distribution) generated by a metallic conical tip (30$^{\circ}$
aperture, 25 $\mu$m height) on a flat graphite film.  Tip-film
potential difference 150 V, distance $d$ = 500 nm.}
\end{figure}
\begin{figure}
\includegraphics[width=8.5cm]{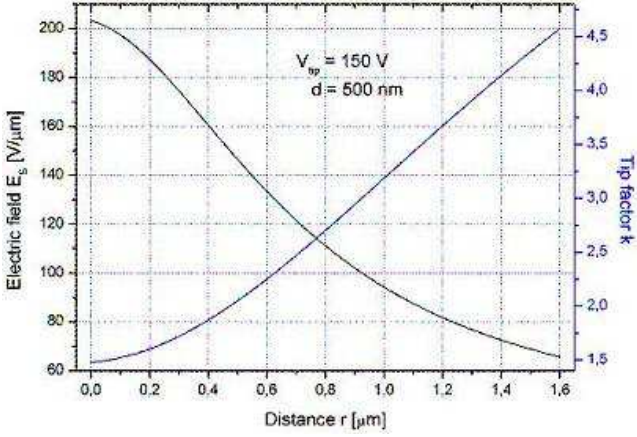}
\caption{\label{fig:epsart} Electric field magnitude $E_S(r)$ on the
graphite film surface and tip correction factor $k(r)$ as a
function of the horizontal distance $r$. Voltage bias 150 V,
tip-flat surface distance 500 nm.}
\end{figure}
As expected, the field is perpendicular to the graphite surface
and its modulus $E_S(r)$  is a decreasing function of the distance
$r$. $E_S(r)$   is plotted in Fig. 6 together with a tip
correction factor $k(r)$ defined as
$E_S(r)=\frac{V}{d}\frac{1}{k(r)}$. Notice that the electric field
on the surface of the apex of a tip, with curvature radius $\rho$,
is often expressed as $E_{Tip}=\beta V$, where
$\beta=[k(0)\rho]^{-1}$ is called the field enhancement factor of
the tip. The correction factor $k$, that we have introduced,
depends on the geometry of the tip and is independent of the bias
voltage V.

A numerical calculation based on the image charge method  (see
reference \cite{campo1, campo2}) for a hyperbolic tip similar to
ours gives a value close to our $k(0) = 1.48$. By using $E_S(r)$
in combination with the FN formula (1), and, thanks to the radial
symmetry, by dividing the emitting surface in concentric annuli
with center P, we obtained the predicted behaviour of the FE
current. In the numerical calculation,  we assumed $\Phi = 4.8$ eV
as workfunction of the CNTs \cite{phi}, while a constant field
enhancement $\gamma_{eff}$ = 30 was used to reproduce our
experimental data since with a voltage bias of 150 V and at $d$ =
500 nm we measured a current of about 10$^{-5}$ A.

Fig. 7a shows that more than 99 \% of the current is emitted from
a circle of radius $r\leq d$, on which the field $E_S(r)$  is
reduced by $\sim$38\% (while $k$ varies from 1.48 to 2.0 (Fig.
7c)). The exponential behaviour implies that a modest variation of
the field has a drastic effect on the value of the emission
current.

Fig. 7b shows that maximum emission occurs from annuli of radius
$r$ $\sim$ $d/3$ (thickness dr = 0.5 nm): field emission is
initially dominated by the area increase, that, for $r\geq d/3$ ,
is overwhelmed  by the fall of the current density with increasing
$r$.

Relative emission (i.e. percentage of the total) from a certain
zone strongly depends on the applied voltage. To quantify this
dependence, we report in Fig. 8 the fraction of the current, as a
function of the bias voltage, that occurs from three circles,
centered on P, of radius  $d$/2, 2$d$/3 and $d$, respectively. We
can conclude that a circle of radius $r \sim d$ is a very good
estimation of emitting area over our entire sweeping voltage,
while $r = 2d/3$ can be a sufficient value for applied voltages
less than 100 V.

The calculated FE current versus the bias voltage and its
corresponding  FN plot are shown in Fig. 9. Remarkably,  the FN
plot is well fitted by a straight line, an expected result since
\begin{eqnarray}
 I&&=\int_0^{-\infty}dr 2 \pi r j(r)\approx \nonumber \\\approx
 &&\int_{0}^{r_{eff}}dr 2 \pi
 r \frac{a\gamma ^2}{[dk(r)]^2\Phi}V^2 exp[-\frac{bd\Phi^{3/2}k(r)}{\gamma}
 \frac{1}{V}]
\end{eqnarray}
(with $r_{eff}\leq d$ ) makes I/V$^2$  an exponential function of
1/V.

For $0 \leq r \leq r_{eff}$, we can neglect the variation of k(r)
and give it a constant effective value $k_{eff}$. Hence
\begin{equation}
 I \approx \pi r^2_{eff} \frac{a \gamma^2_{eff}}{[dk_{eff}]^2
 \Phi}V^2 exp\left[-\frac{b\Phi^{3/2}d\cdot k_{eff}}{\gamma_{eff}\cdot }\frac{1}{V}\right]
\end{equation}

By using our simulated data, we evaluate $r_{eff}$  and $k_{eff}$
from the slope $m$ and the interception $y_0$ of the FN plot,
which result respectively:

$k_{eff}=\frac{m\gamma}{b\Phi^{3/2}d}\approx$ 1.6 and
$r_{eff}=\frac{m e^{y_0/2}}{\sqrt{\pi a}b\Phi} \approx$  300 nm.

As a conclusion, we can analyze our experimental data using
formula (4) with $k_{eff}\approx$ 1.6  and $r_{eff}\approx
\frac{3}{5}d$.

\begin{figure}
\includegraphics[width=8.0cm]{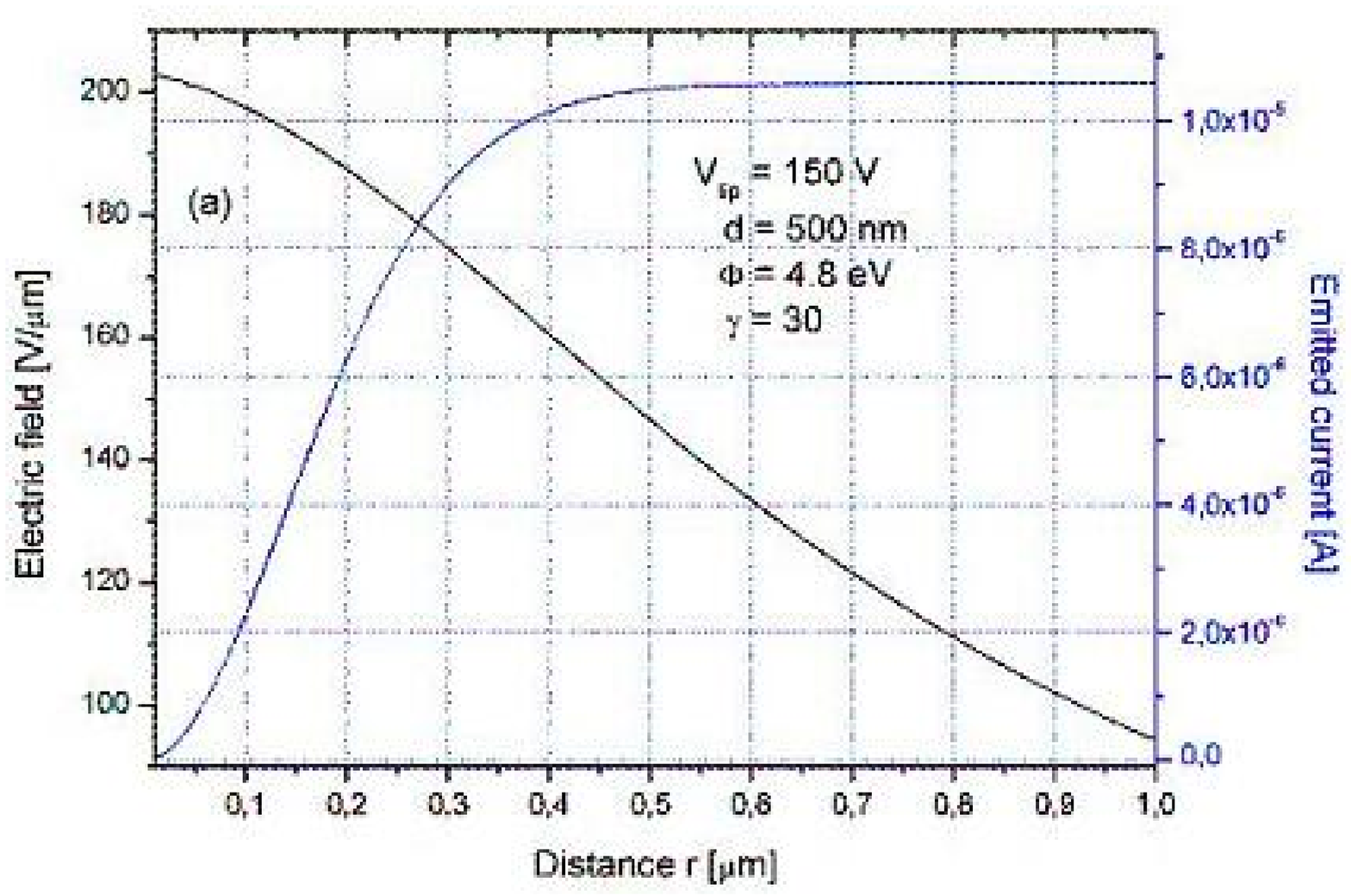}

\includegraphics[width=8.0cm]{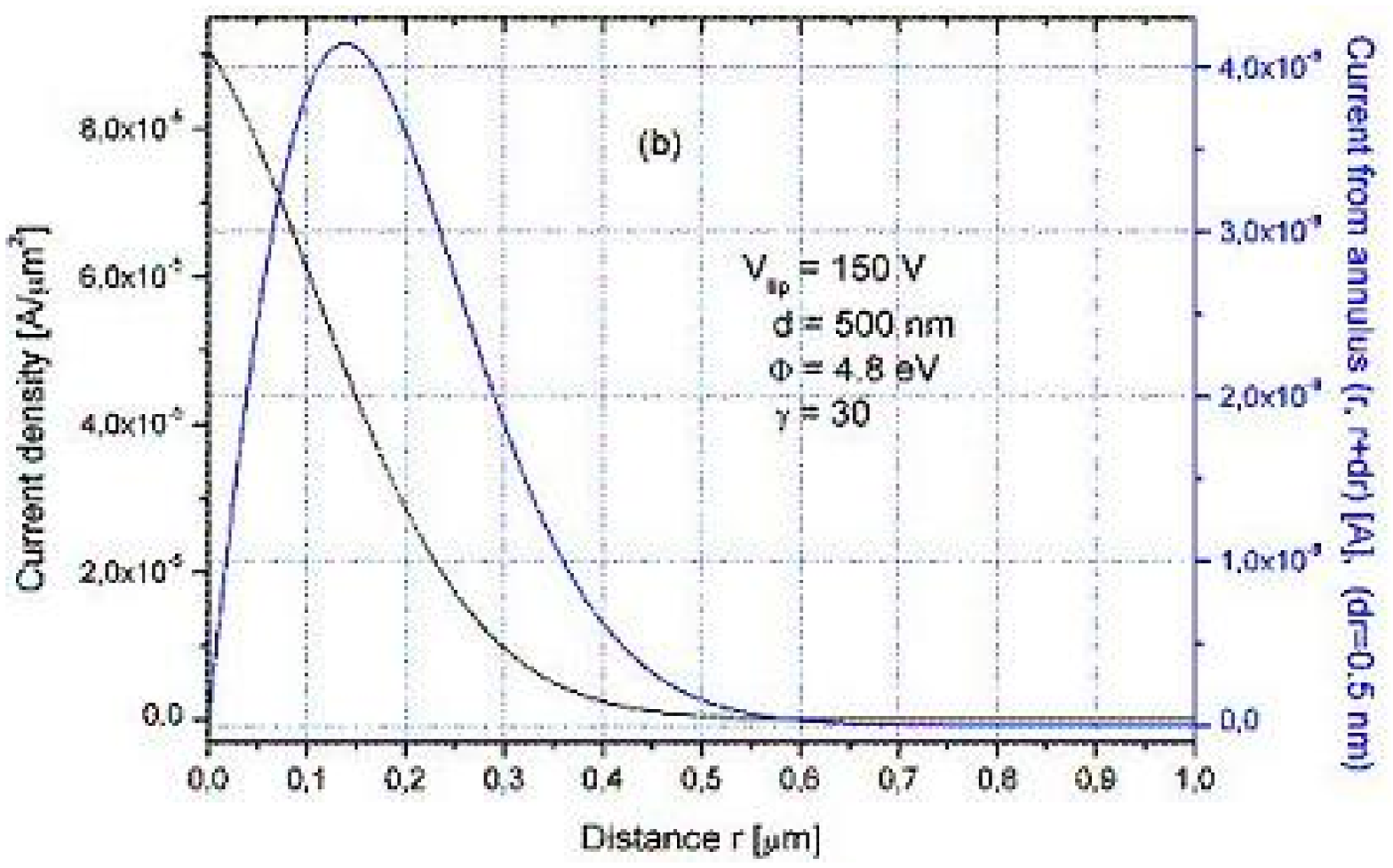}

\includegraphics[width=8.0cm]{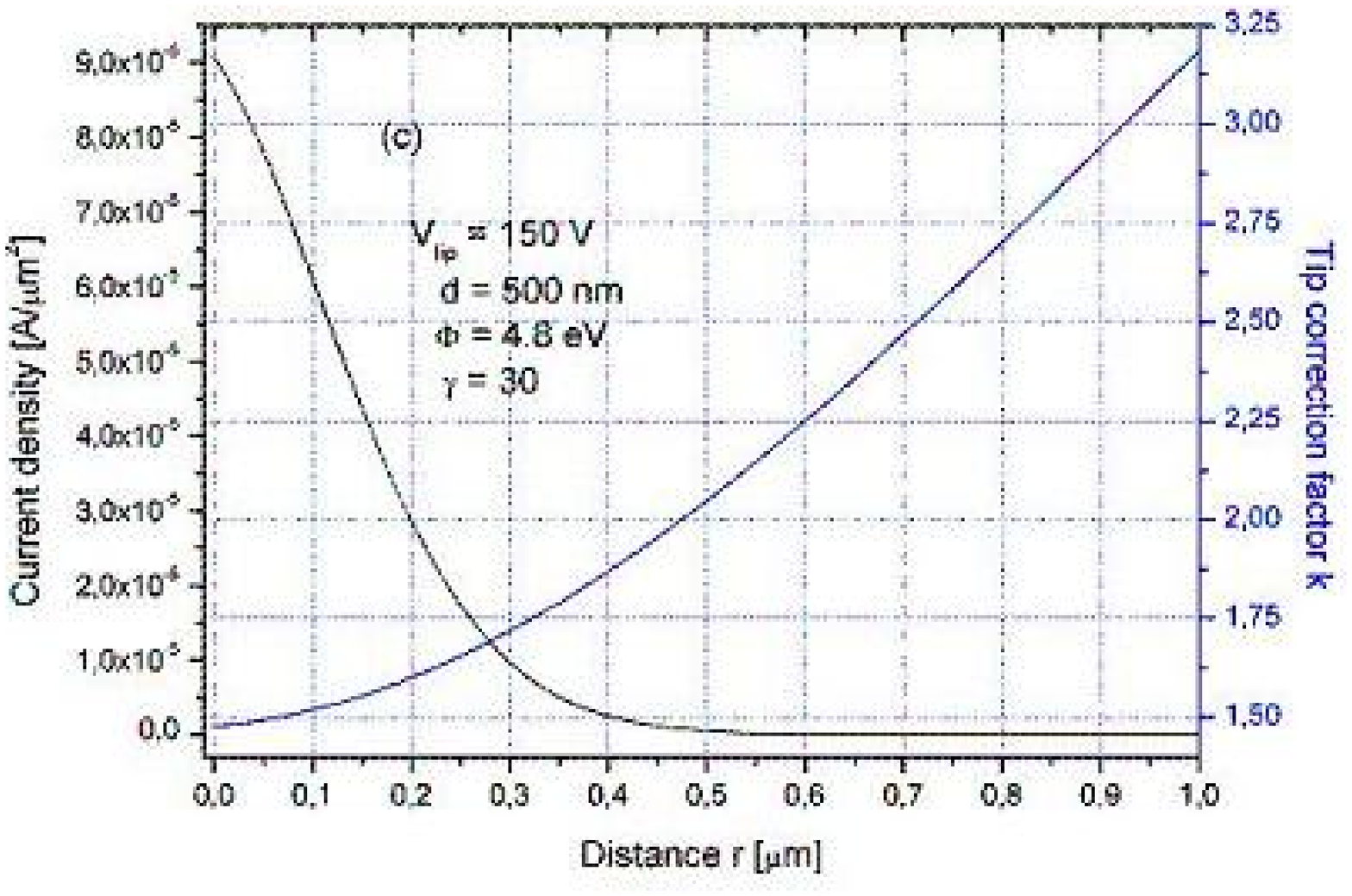}
\caption{\label{fig:epsart} Predicted FE current $I$, density of
current J and current d$I$ from annulus ($r$,$r$+d$r$) as a
function of the horizontal distance $r$.  dr=0.5 nm in the
numerical calculations.}
\end{figure}

\section{Experimental results and discussion}

\subsection{I-V curves and FN plots}

An example of a current-voltage characteristic measured by our
apparatus is shown in Figures 10a-c. These data were obtained at a
pressure of $\sim10^{-4}$mbar in non-contact mode by using an
Au-coated AFM tip at a distance $d\approx$800 nm from the CNT film
surface and for an applied voltage stepping in the range (-210,
+210) V.

Above the sensitivity limit of our SMU ($\sim$1 nA at 100 V), we
can observe a rapid rise of the current with the absolute value of
the applied voltage. Indeed, electrons are extracted from the CNT
film at voltages $\geq$ +120 V, while for negative biases below
-140 V, the current is field emitted from the AFM tip; .

\begin{figure}
\includegraphics[width=8.0cm]{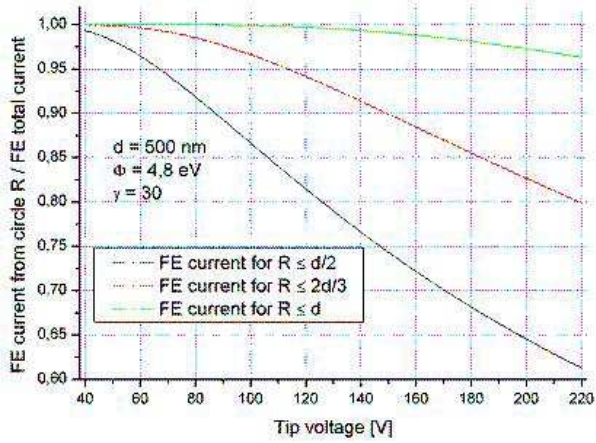}
\caption{\label{fig:epsart} Percentage of the FE current from
circles of radius R=$d/2$, 2$d$/3 and $d$ and centered on P as a
function of the bias voltage.}
\end{figure}
\begin{figure}
\includegraphics[width=8.0cm]{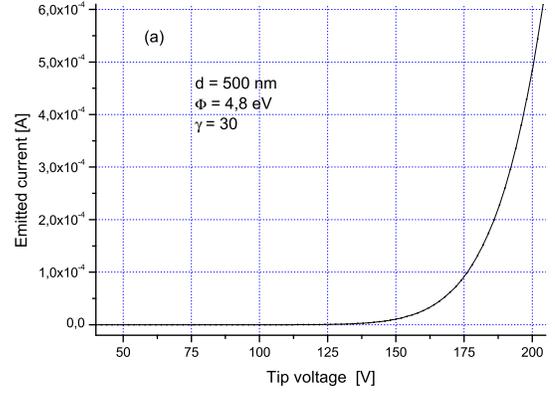}

\includegraphics[width=8.0cm]{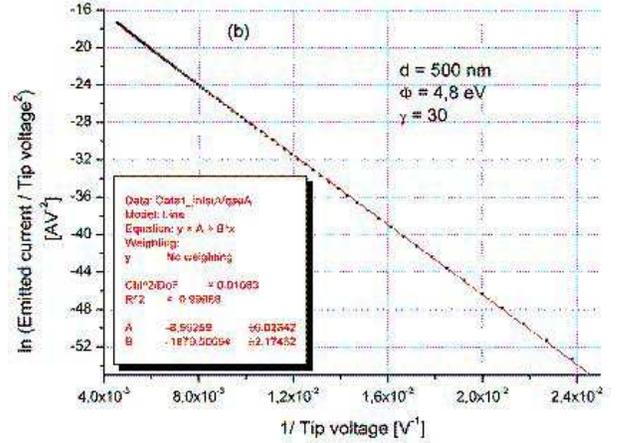}
\caption{\label{fig:epsart} Predicted FE current vs bias voltage
and relative FN plot.}
\end{figure}

The FE current from the CNTs occurs at a lower macroscopic
electric field, is significantly larger and presents less
fluctuations than that from the AFM tip, confirming the CNT film
as a higher quality emitter. A maximum current of $4.5\cdot
10^{-5}$A, corresponding to a current density $2-6\cdot 10^{3}$
A/cm$^{2}$, is achieved without emitter failures. For the two sets
of data (FE from CNT and tip), the different slopes
($m=\frac{k_{eff}b\Phi^{3/2}d}{\gamma}$) of the straight lines in
the FN plot (Fig. 10c) can be attributed to a higher CNT field
enhancement factor (gold has a workfunction of 5.28 eV, higher
than that supposed for a MWCNT).

\begin{figure}
\includegraphics[width=8.0cm]{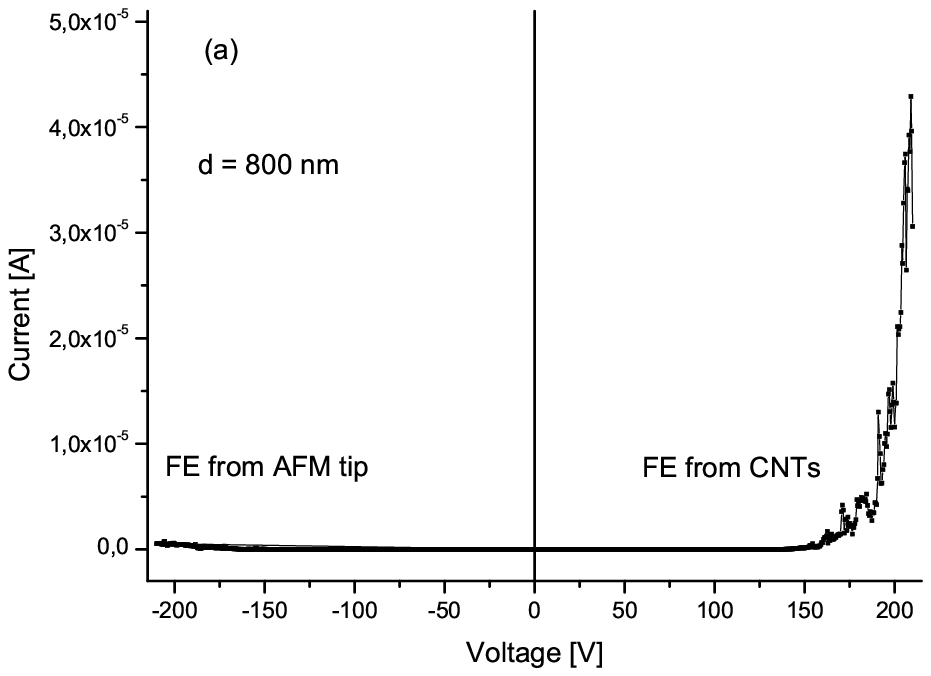}

\includegraphics[width=7.5cm]{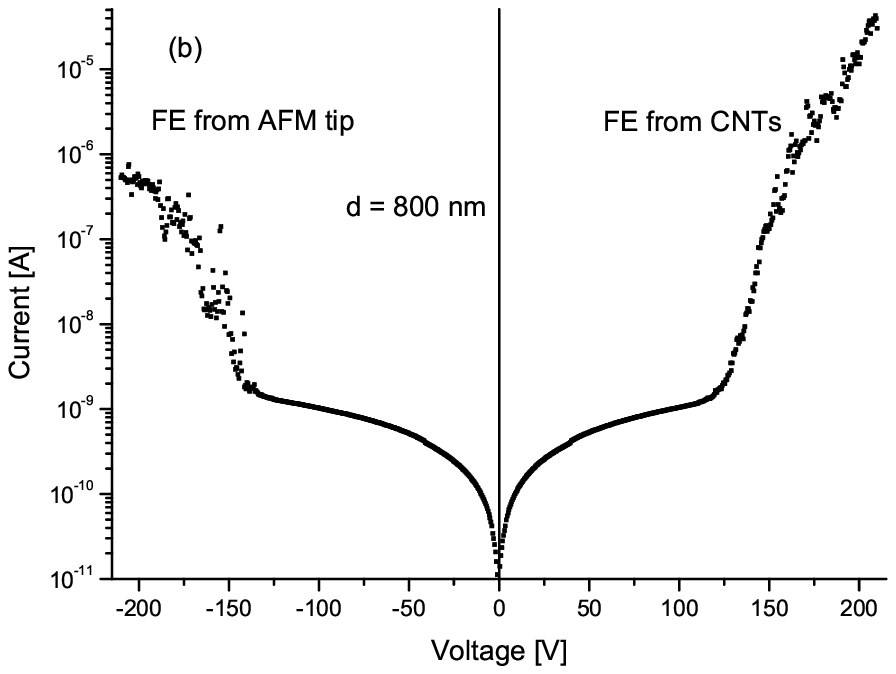}

\includegraphics[width=8.0cm]{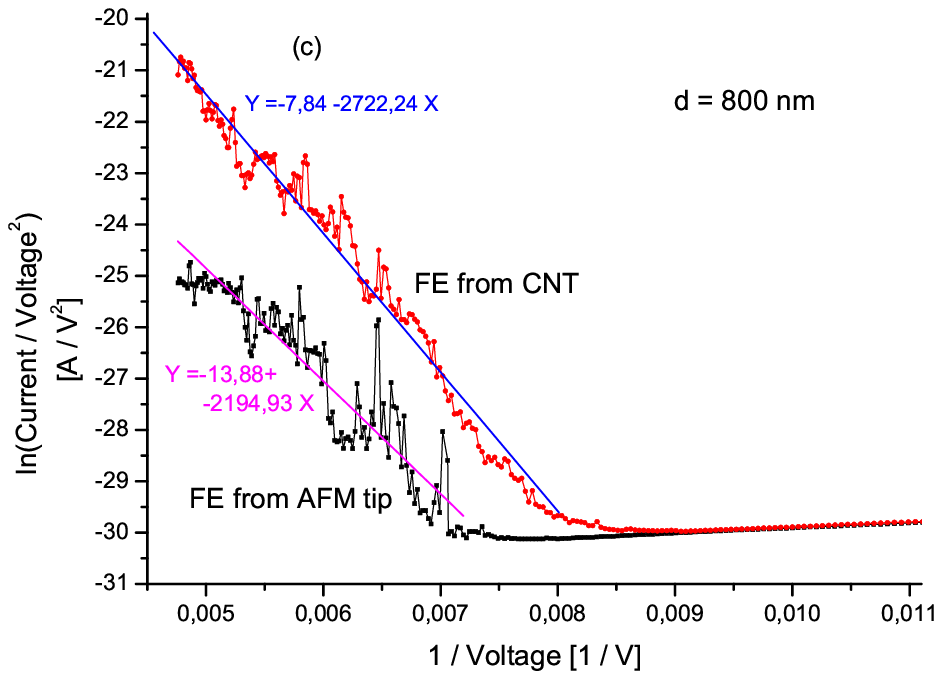}
\caption{\label{fig:epsart} Field emission current versus bias
voltage in (a) linear and (b) logarithmic scale. (c)
Fowler-Nordheim plot. Tip-CNT distance $d\approx$  800 nm,
pressure $10^{-4}$ mbar, room temperature.}
\end{figure}

\begin{figure}
\includegraphics[width=6.0cm]{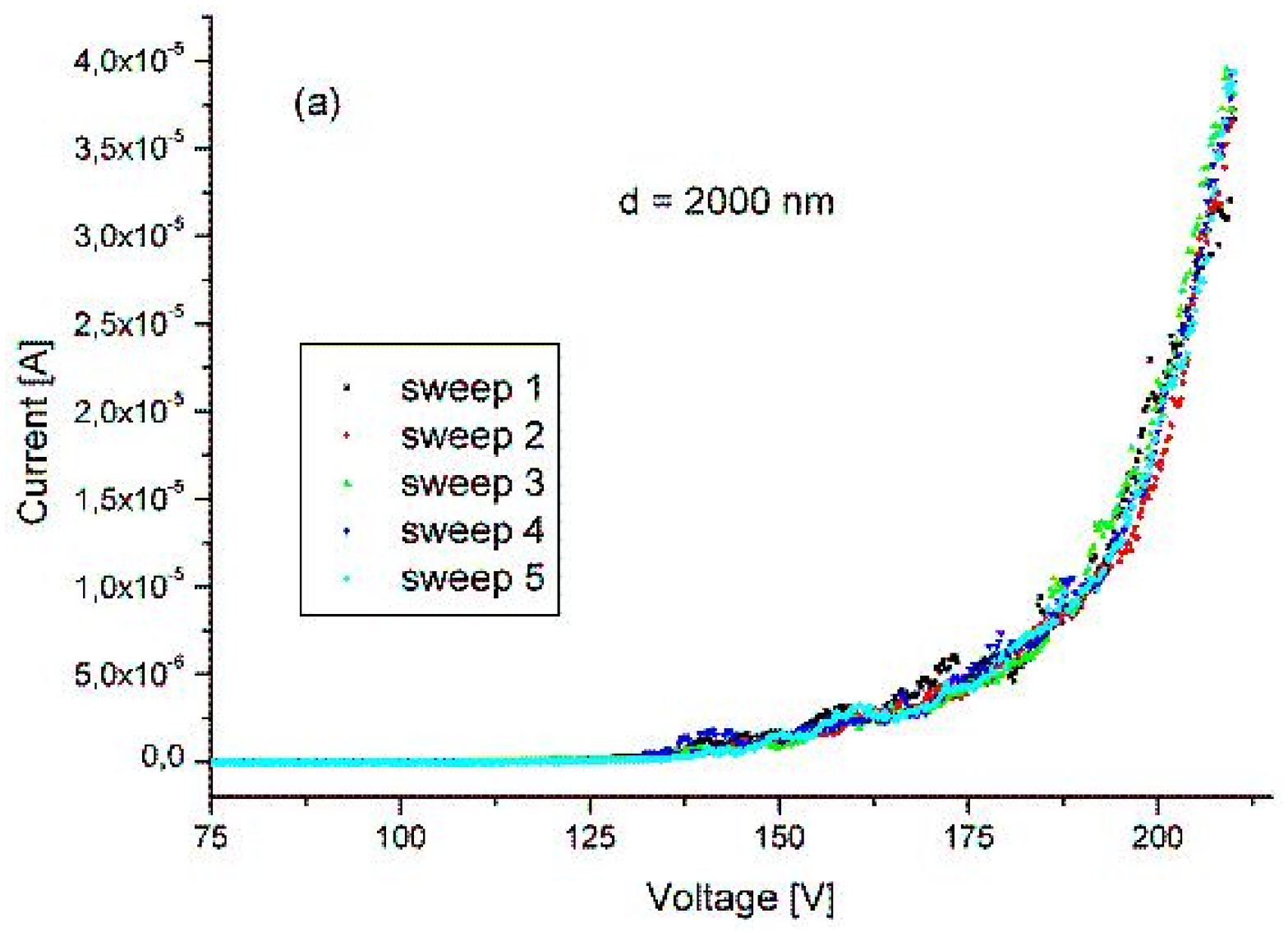}

\includegraphics[width=6.0cm]{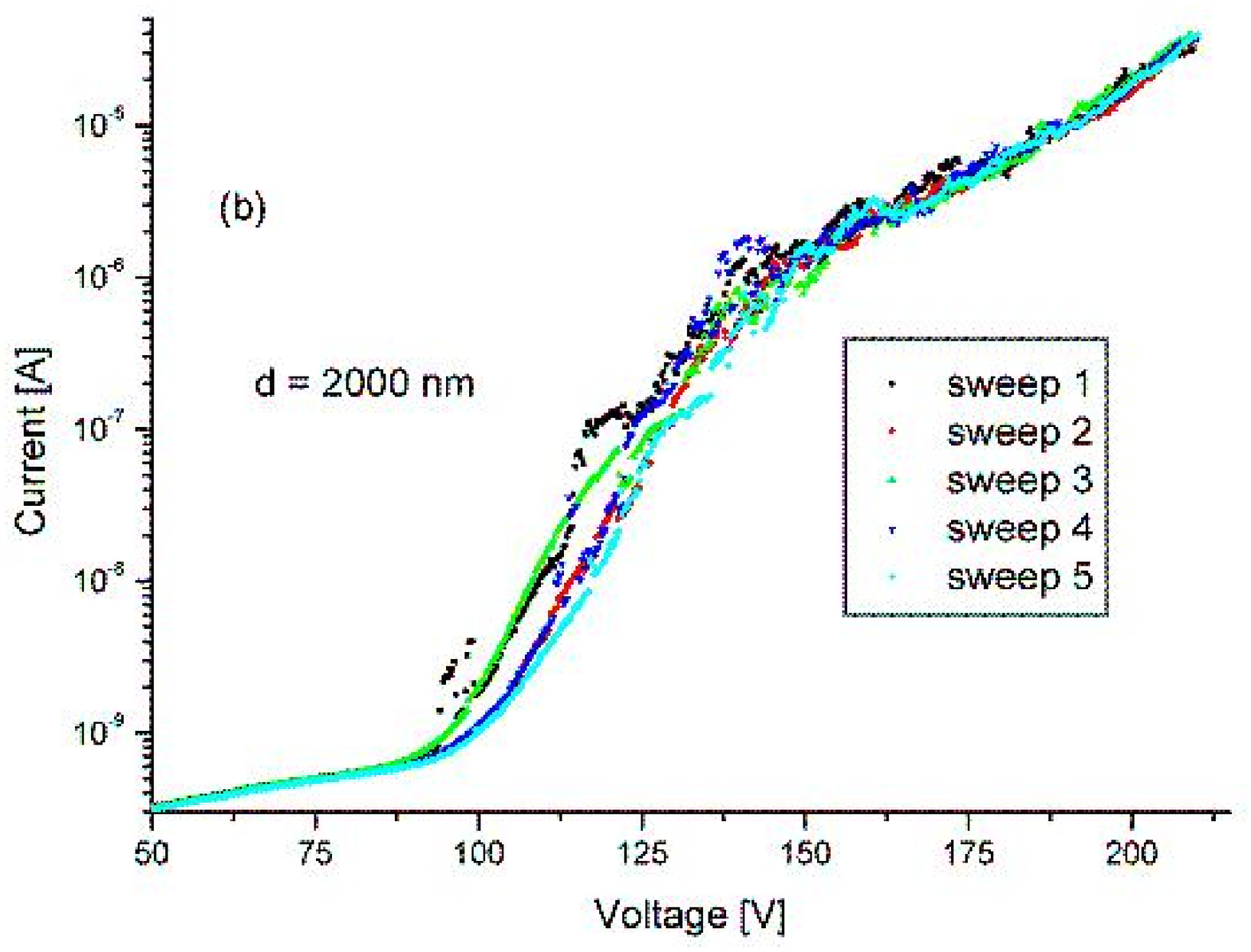}

\includegraphics[width=6.0cm]{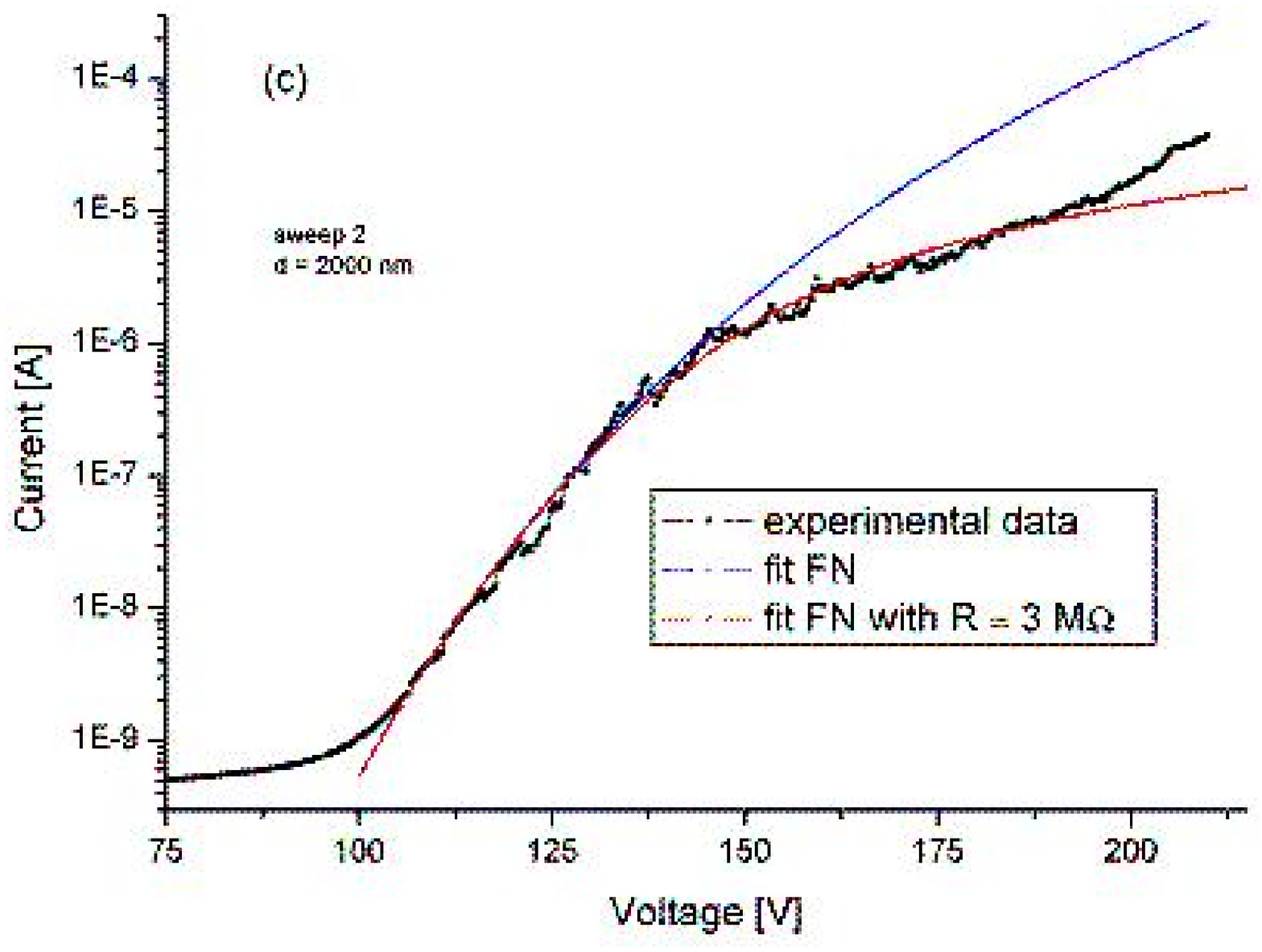}

\includegraphics[width=6.0cm]{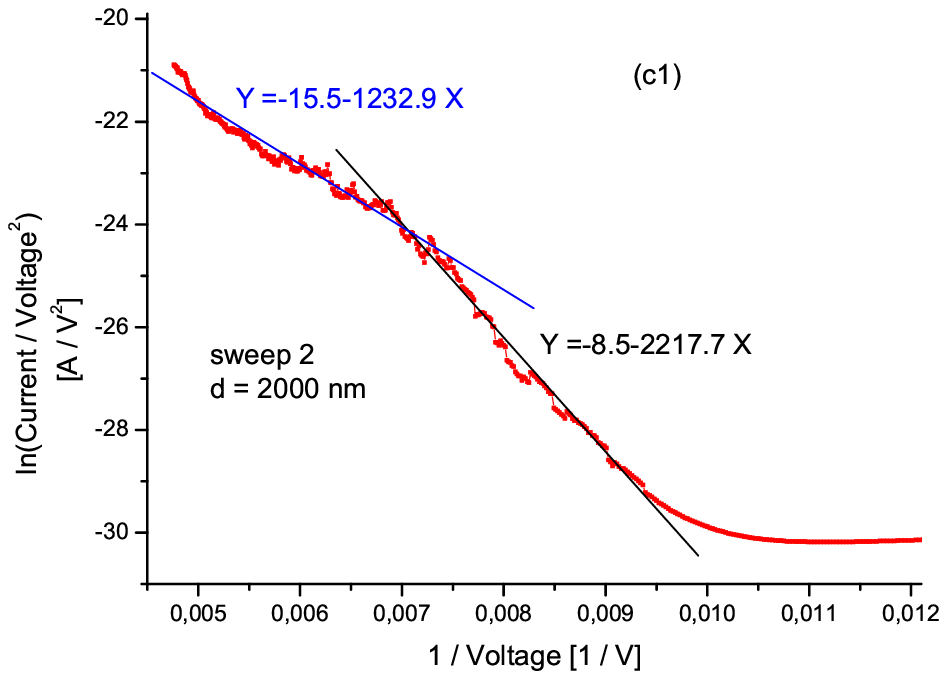}
\caption{\label{fig:epsart} Field emission current versus bias
voltage in (a) linear and (b) logarithmic scale. (c) Current with FN and resistance modified FN fitting and
(c1) Fowler-Nordheim plot for sweep 2. Tip-CNT distance $d\approx$
2 µm, pressure $\sim$ 10$^{-8}$ mbar, room temperature.}
\end{figure}

The distance $d$ is measured from the approach point on the CNT
film surface in AFM non-contact mode. The small force (1-10nN)
applied by the tip to the surface in this modality likely does not
change the local CNT topography by inducing displacements or
vibrations of the highly moving and the flexible nanotubes. As
already mentioned, the CNT topography and therefore the distance d
can be affected by the application of the external voltage, which
attracts the CNTs towards the tip. Consequently, the approach
point is sometimes non reproducible and an accurate measurement of
d is impossible. Furthermore Joule heating at high currents can
modify the CNTs (sublimation of the walls, split off, severing,
etc.) and again the distance d can vary. In what follows, we will
consider the parameter d as merely indicative, and when possible,
we will avoid using it in our quantitative analysis.

Assuming $\Phi$ = 4.8 eV, from Fig. 10 c, we calculate $\gamma
\approx$24 and the microscopic field at which observable emission
begins, E$_{turn-on}^{microscopic} \approx$ 2.2 kV/$\mu$m.

A similar measurement, on the same sample, at high vacuum, $\sim
10^{-8}$ mbar, and with $d \approx$ 2 $\mu$m is shown in Fig. 11.
After an electrical stress, consisting in a few voltage sweeps,
necessary to stabilize the emission (see following), a series of 5
sweeps from 0 to 210 V was performed. Some negligible
sweep-to-sweep variations were found, while, with respect to the
measurements at lower vacuum, reduced single sweep fluctuations
were observed.

The FN plot of Fig. 11 c1, for clarity showing only sweep 2,
reveals an interesting deviation from linearity. A change of slope
(knee) appears at a bias of $\sim$145 V (Fig. 11c) and corresponds
to a sort of current saturation above 1 $\mu$A (a closer look at
Fig. 10 c shows a similar, but less pronounced knee, partially
masked by the higher fluctuation level).

Saturation has been observed on individual multi-walled and
single-walled nanotubes and to a lesser extend on nanotube films,
and different explanations have been proposed
\cite{saturation1,saturation2}. We ascribe this to a reduction of
the applied field caused by the voltage drop on the series
resistance R of the circuit model in Fig. 3.

Assuming R constant, formula (4) is modified as

\begin{equation}
 I = c_1 (V-RI)^2 exp\left(-\frac{c_2}{V-RI}\right)
\end{equation}

The constants $c_1$ and $c_2$, which include $\gamma_{eff}$,
$r_{eff}$ and $\Phi$, can be evaluated by fitting eq. (5) to the
low emission current part of the experimental data, with R = 0.
The result is shown in Fig. 11 c. In the chosen voltage range,
105-145 V, there is a very good agreement with the basic FN
model.

Including the effect of the series resistance is not trivial.
Equation (5) is recursive and has to be evaluated numerically,
with R as fitting parameter. By numerical calculations, we were
able to estimate a $R\approx 3.25 M\Omega$ and obtain the I-V
characteristic shown in Fig. 11 c. With such R, Eq.(5) constitutes a good fitting to our data up to a voltage of
about 190V; a rapid rise of the current is observed afterwards,
likely due to additional nanotubes that enter the FN regime at
such high electric fields.

Possible origins for the resistor limited emission regime can be
either a high internal resistance of the CNT, due to defects or
heating for example, and/or to a resistive path to the SMU
(resistance tube-to-tube, tube-to-substrate, CNT-to-metal and substrate-to-metal contact, etc.).

From the value of the emission current and of the voltage drop one
can estimate the power dissipated on R, which is up to few mW in
our measurements. This power is dissipated on a very small volume
and can be sufficient to cause contact melting, detachment from
the substrate, CNT severing, etc., with consequent emission
degradation. An example of an event of high current degradation
occurring at  $I\approx 30 \mu A$ is shown in figure 12.
\begin{figure}[t]
\includegraphics[width=8.0cm]{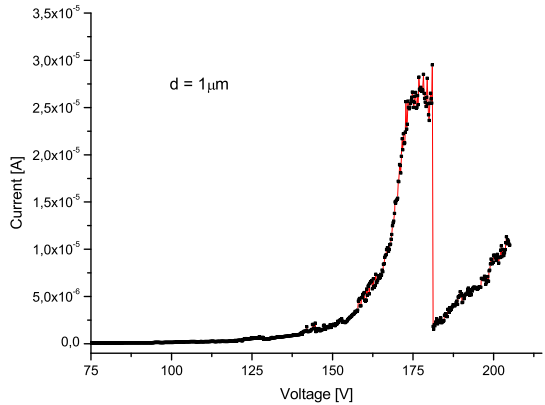}
\caption{\label{fig:epsart} I-V characteristic showing an event of
high current degradation occurring at V $\approx$ 180 V and I
$\approx$ 30 $\mu$A. Probably the emission after the degradation
do not originate from the same tubes as before but from tubes
nearby, which were before concealed by the original emitters.}
\end{figure}

Finally, taking into account all the 5 sweeps, from the lower
current part of the FN plot, a field enhancement factor $\gamma
\approx 60-70$  and a E$_{turn-on}^{microscopic} \approx$ 2.0-2.2 kV/$\mu$m can be evaluated.

\begin{figure}
\includegraphics[width=7.0cm]{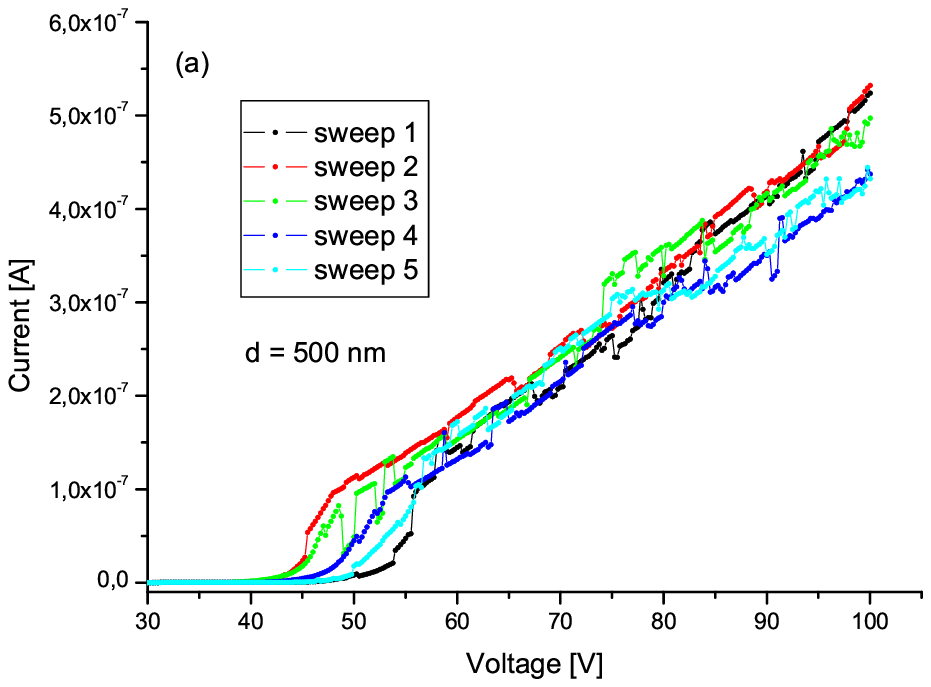}

\includegraphics[width=7.0cm]{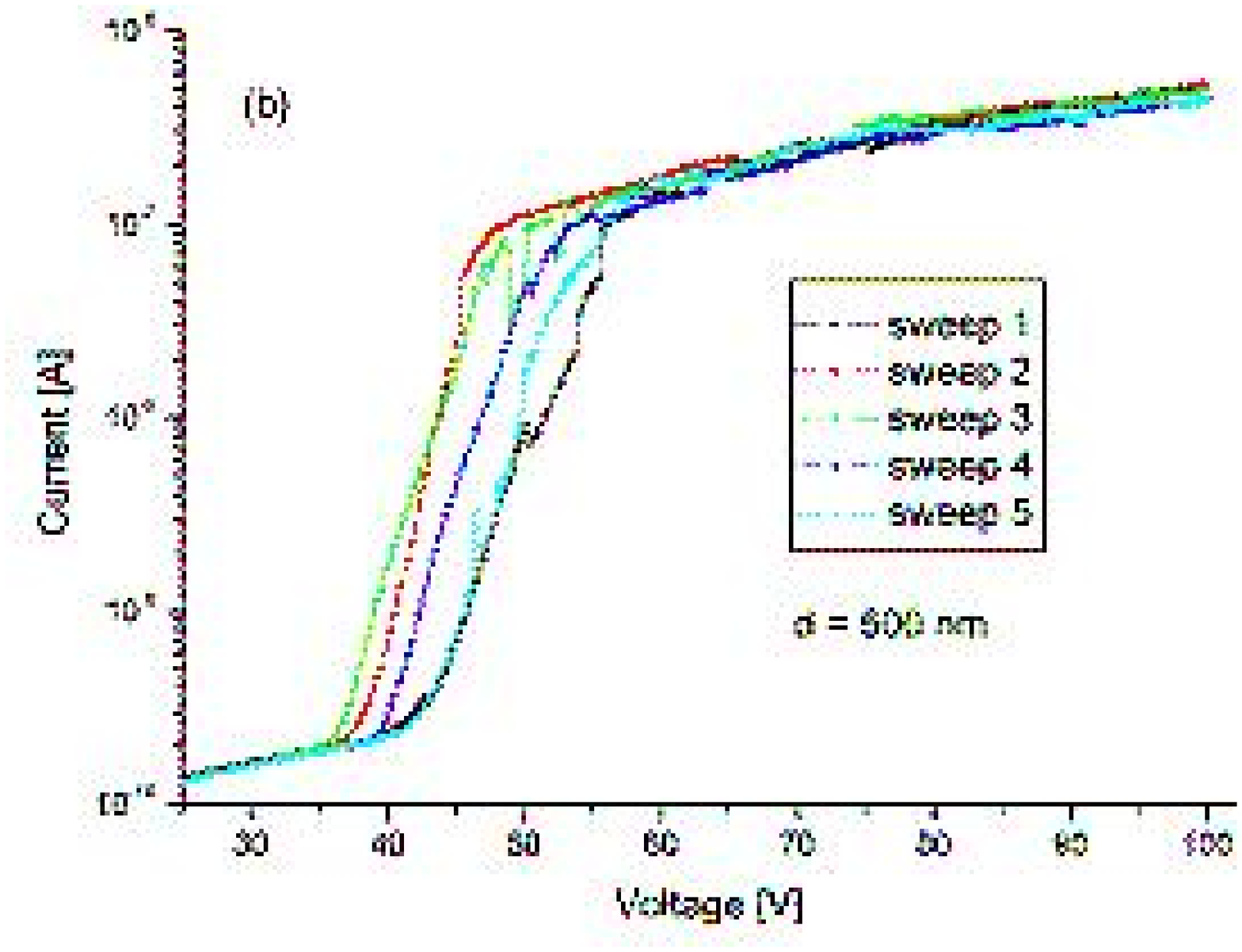}

\includegraphics[width=7.0cm]{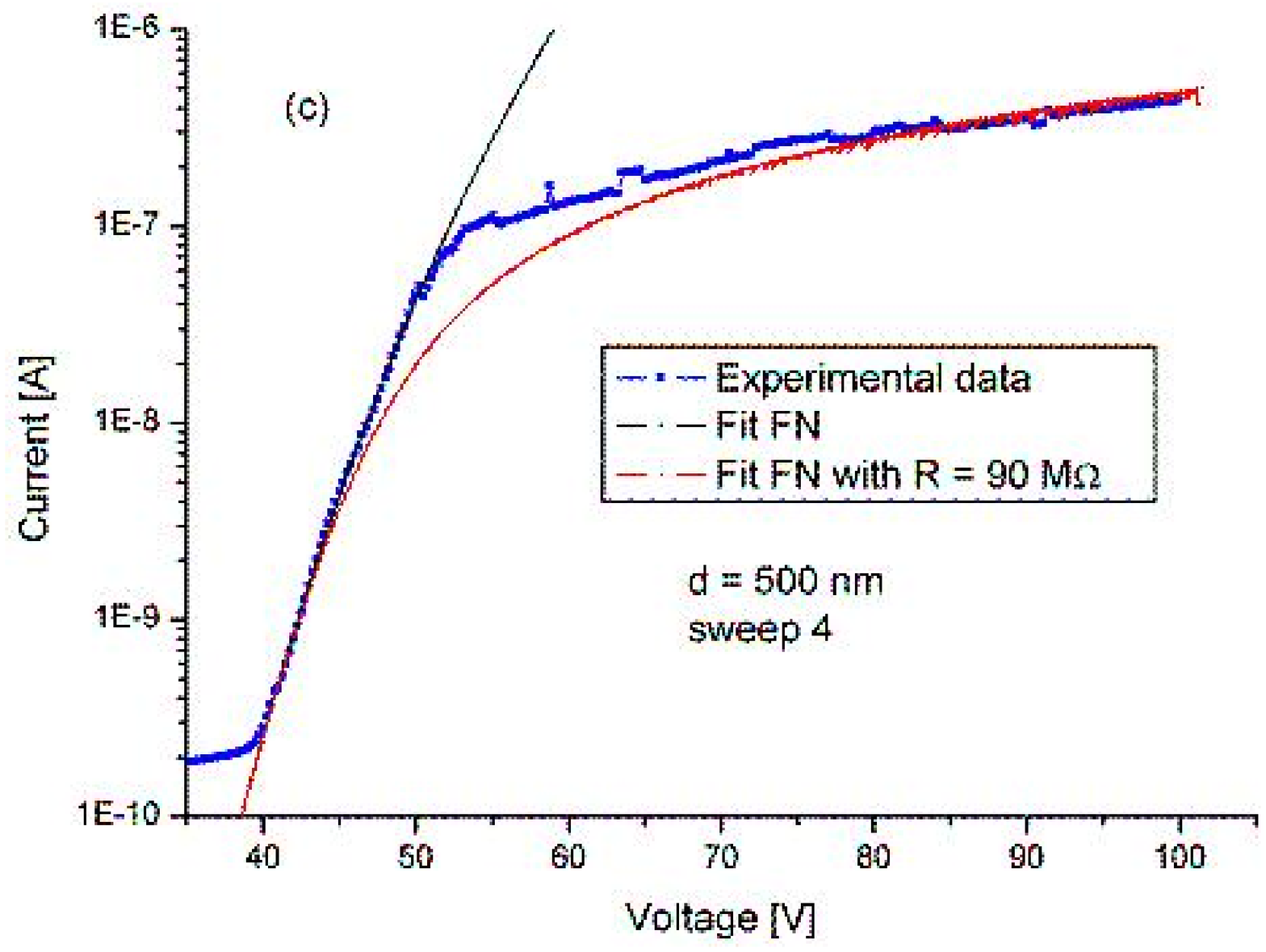}

\caption{\label{fig:epsart} Characteristics I-V with I in linear
(a) and logarithmic scale (b,c). After a first steep rise (low
current regime), a resistor limited emission regime (saturation)
with R $\sim$ 80 M$\Omega$ take place. Pressure 10$^{-8}$ mbar,
$d\approx$500 nm. Room temperature.}
\end{figure}

The series resistance R can be expected to change across the film.
Fig. 13 shows the I-V characteristics, with the applied voltage
limited to 0-100 V to avoid degradation, where the effect of R,
evaluated around 80 M$\Omega$, is very pronounced. Actually we
face a different phenomenon. At a current of  $\sim$0.1$\mu$A,
corresponding to a voltage of 50 V, a change of conduction
mechanism occurs in the I-V characteristic: FE regime is followed
by a pure ohmic increase of the current. A series resistance
corrected FN model barely reproduces the I-V curve (Fig. 13 c). By
referring to our model circuit, above 0.1 $\mu$A, the field
emission device seems to be replaced by a high value resistor.
This seems to indicate that we were measuring in a region poor of
CNTs or with CNTs damaged by previous electric stress and with
lowered FE capability.  This hypothesis is supported by the
observation, through SEM analysis, of small spots of completely
removed or simply shortened CNTs after uncontrolled electric
stress. Fig. 14 shows zone (dark spot) were CNTs were completely
removed by the passage of uncontrolled current during several
attempts of setting the right parameters for the measurement of
field emission maps (see following).

\begin{figure}
\includegraphics[width=7.0cm]{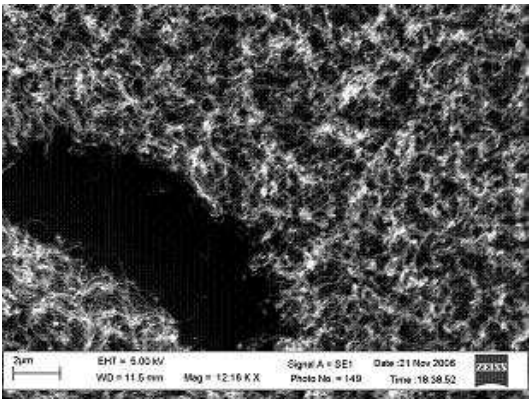}
\caption{\label{fig:epsart} Region of the film with CNTs pulled up
or destroyed by electrical stress.}
\end{figure}

\begin{figure}
\includegraphics[width=7.5cm]{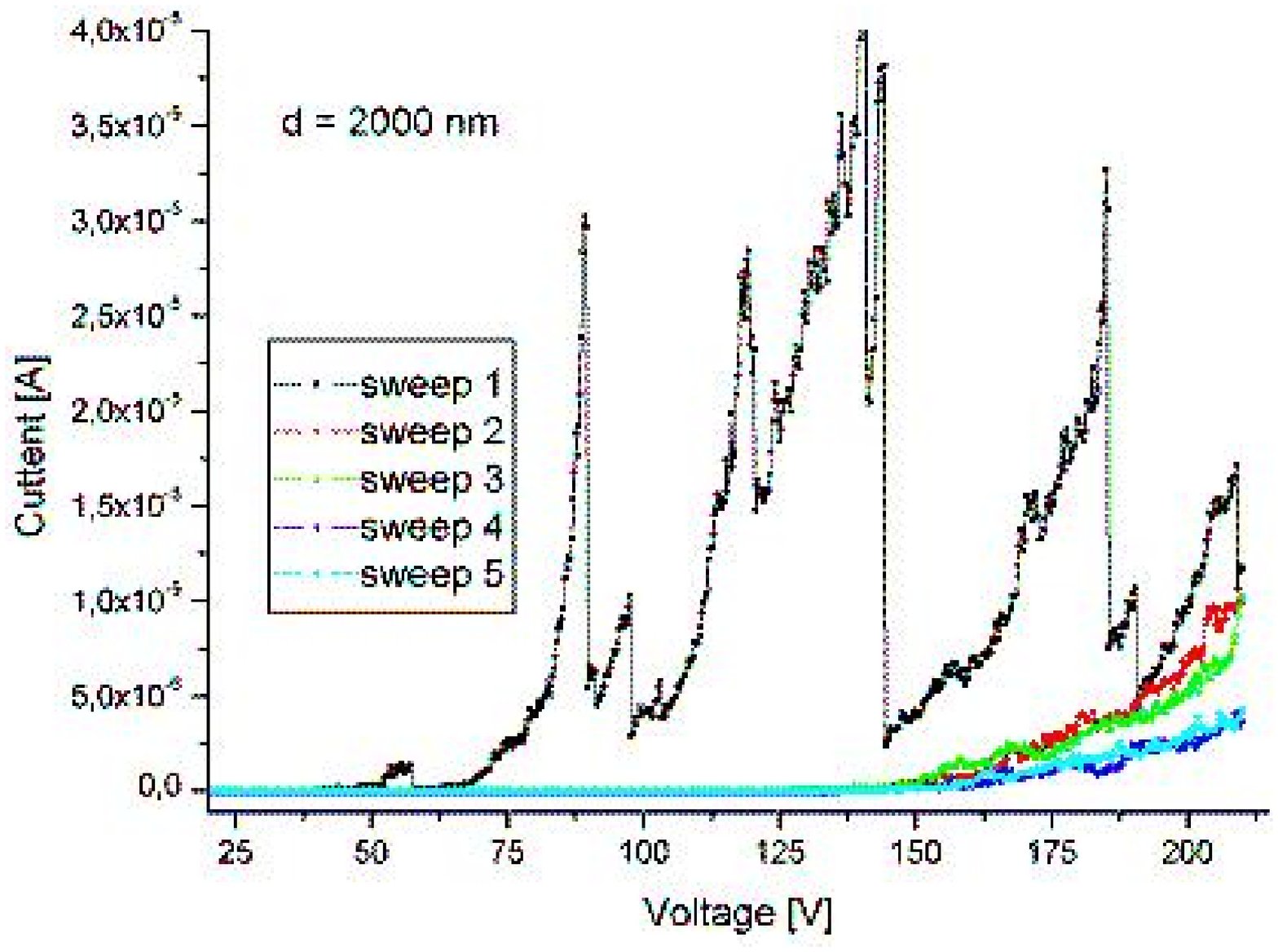}

\includegraphics[width=7.5cm]{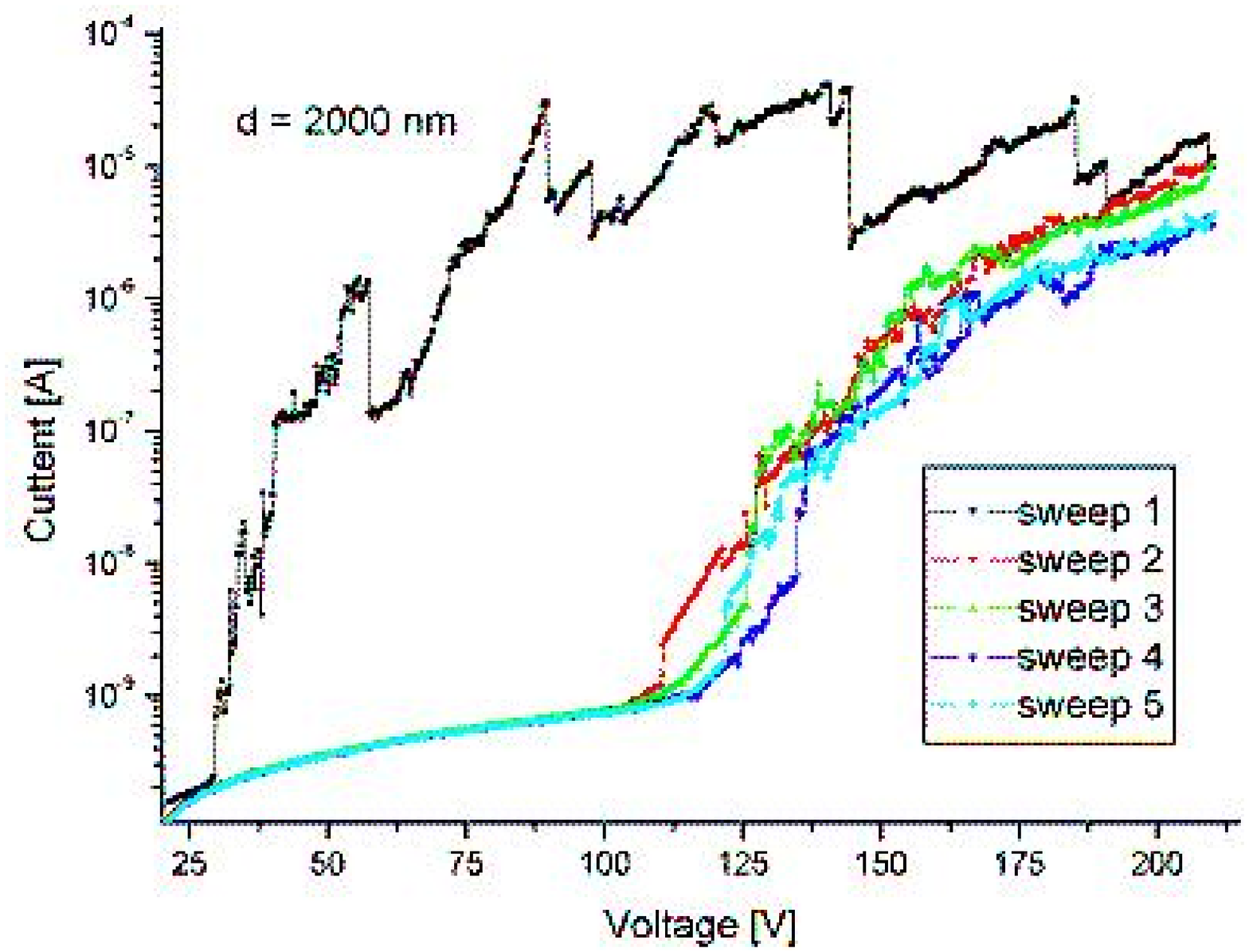}
\caption{\label{fig:epsart} Electrical conditioning showing the
stabilization of the emission current by electric stress.}
\end{figure}

\subsection{Electrical stabilization}

We systematically observed that the initial electrical sweeps on
virgin zones have a positive conditioning effect: irreversible
changes, in fact resulting in a stabilization, were found on the
I-V characteristics.

An example is shown in Fig. 15. During the first sweep, a sudden
rise of the current is observed around 30 V; after entering the
$\mu$A range, the current suffers stepwise -up to one order of
magnitude- drops and, at 210 V, its value is $\sim$17 $\mu$A.
Following sweeps show a completely different behaviour, with
higher turn-on voltage, 110-115 V, and considerably lower current
(the higher the voltage the lower the current difference). For
sweeps 2 and 3, at 210 V,  a current of 10 $\mu$A is measured;
such current ($\sim$4 $\mu$A) is halved during sweeps 4 and 5.

We can attribute the higher current of sweep 1 to a single (or a
few) longer nanotube(s) with dominant FE; a current around 1
$\mu$A gradually degrades such tube(s), till a complete
destruction. After that FE becomes more stable.

Sometimes the stabilization process requires  more sweeps.

A second example is shown in Fig. 16, where it is emphasized how
the FN plot is evolving toward a straight line after an electrical
conditioning. Apart fluctuations, the slope of the FN plot  for
sweep 1 and for the successive sweeps are comparable. This
suggests that the field enhancement factor and the workfunction in
both cases remain the same.

The slope of the FN plot, obtained from a fit of sweep 2, 3 and 4
together, can be used as usual to calculate the field enhancement
factor, which results $\gamma \approx$23.

In Fig. 16 b and c, the superposed (magenta) line refers to the
prediction of the simulation based on a simple FN model (eq. 3),
previously discussed. A little discrepancy of simulated and experimental
data is observed at high current, since the effect of the series
resistance was not included in the simulation.

In addition to destruction of thinner and longer nanotubes, other
mechanisms involved in the electrical conditioning can be
desorption of adsorbates caused by CNT heating, topography changes
due to CNT stretching and re-orientation, particulate cleaning,
etc.

Adsorbates \cite{adsorbates1,adsorbates2}, as different types of
gases, are always present at the CNT surface, creating
nanoprotrusions i.e regions of reduced workfunction and increased
enhancement factor, where field emission begins at low electric
fields. The formation and the electric field-driven surface
diffusion of those nanoprotrusions can cause the observed
instabilities of the FE current. At large currents, the local
temperature becomes high enough to evaporate some of the
adsorbates, provoking drops in the FE current.
\begin{figure}[h]
\includegraphics[width=5.8cm]{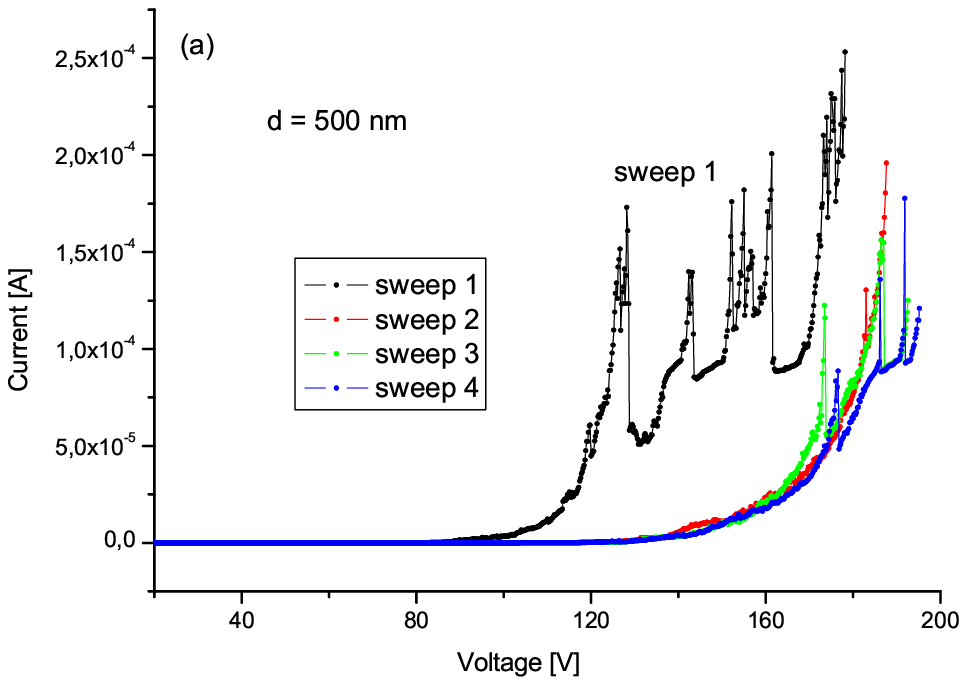}

\includegraphics[width=5.8cm]{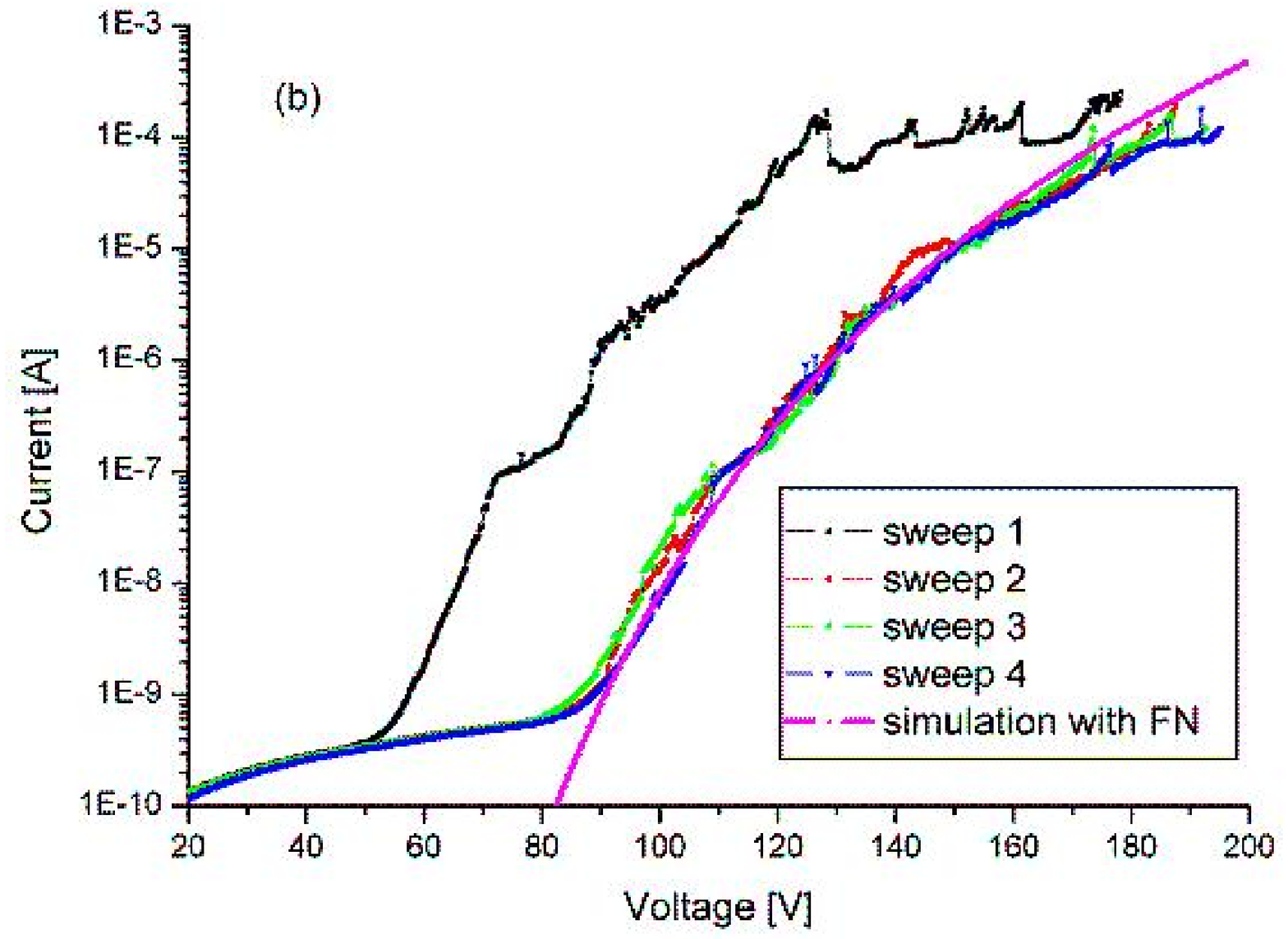}

\includegraphics[width=5.8cm]{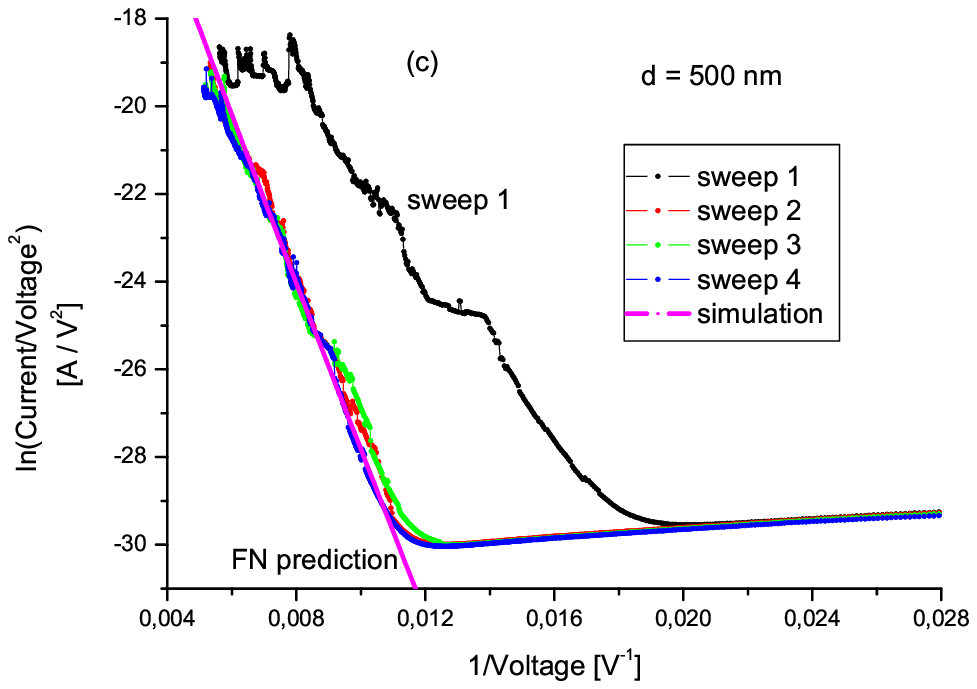}

\includegraphics[width=5.8cm]{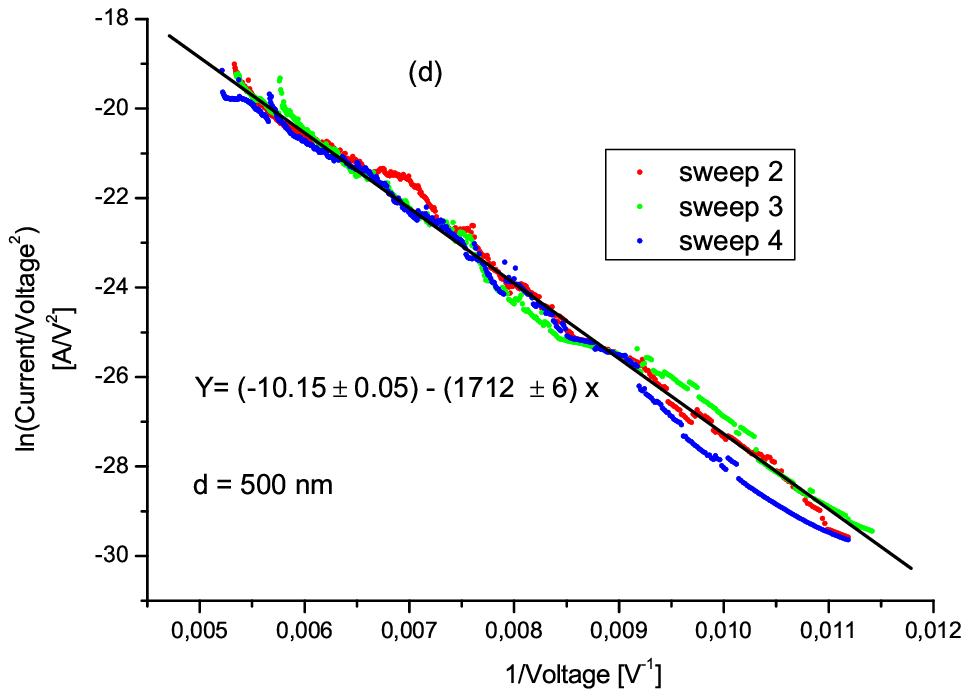}
\caption{\label{fig:epsart} FE for several voltage sweeps.
Instabilities are reduced after first sweep and linearization of
FN plot is obtained. (a) linear and (b) logarithmic current versus
applied voltage, (c) Fowler-Nordheim plot (d) FN plot with linear
fit for sweeps 2,3 and 4. Tip-CNT distance 500 nm. Room
temperature.}
\end{figure}

\subsection{Field enhancement factor}

The field enhancement factor is the typical figure of merit given
to qualify field emitters.  As pointed out in reference
\cite{doubt}, $\gamma$ is strongly dependent on the measurement
setup and a significant comparison of $\gamma$ values is possible
only when measurements are performed under the same experimental
conditions.

\begin{figure}
\includegraphics[width=8.0cm]{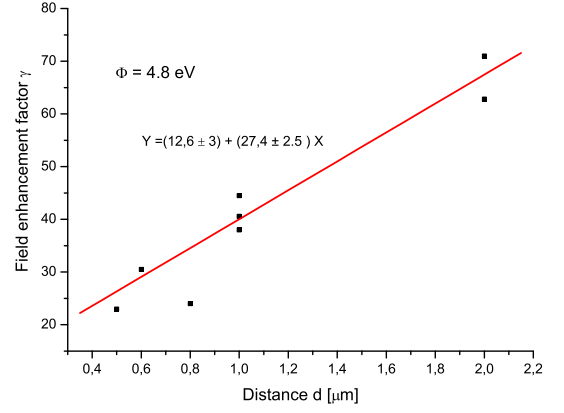}
\caption{\label{fig:epsart} Field enhancement factor as a function
of the inter-electrode distance $d$.}
\end{figure}

In our previous measurements, we evaluated $\gamma$ from the slope
$m$ of the straight line fitting the data in the FN plots, $\gamma
= \frac{b \cdot \Phi^{3/2}\cdot d \cdot k_{eff}}{m} $ assuming
$\Phi$= 4.8 eV and $k_{eff}$ = 1.6.

The field enhancement factor is known to be an increasing function
of the inter-electrode distance and, taking into account our short
d (0.5-2 $\mu$m), we consider to have obtained rather high values
of $\gamma$, likely due to the conspicuous length of our
nanotubes. Published values range from few hundreds to several
thousands, but at inter-electrode distances orders of magnitude
higher than ours \cite{enhancement1,eh2,eh3,eh4}.

Significant field screening effect (the applied electric field on
the apex of a CNT is screened from the neighbouring CNTs
\cite{screening1,scr2}) can be expected  on our sample because of the
high density of CNTs (it has been demonstrated that the optimal
condition for high emission is a tube spacing twice the CNT
length).

To investigate the dependence of the field enhancement factor on
the inter-electrode distance, we estimated $\gamma$ for several
values of $d$. Despite the incertitude on the distance, we found a
monotonic increase of $\gamma$  with $d$ ($\gamma \sim 60-70$ at
2$\mu$m), as shown in Fig. 17. Linear extrapolation to larger
values of $d$ results in higher $\gamma$ values than those found
in Refs. \cite{gamma1,gamma2,gamma3}.

\subsection{MWCNT  workfunction }

The interception of the FN straight line with the y-axis can be
used to evaluate the workfunction $\Phi$ as well:

\begin{equation}
 \Phi=\frac{m e^{y_0/2}}{\sqrt{\pi a}b}\cdot \frac{1}{r_{eff}}
\end{equation}

Formula (6) assumes an accurate knowledge of the emitting area and
of the interception $y_0$  (the exponential factor makes $\Phi$ very
sensitive on  $y_0$). Both those parameters are usually subjected
to great incertitude  and (6) provides a very rough estimation of
$\Phi$. Inversely, given $\Phi$,one can use $y_0$ to evaluate the
effective emitting area \cite{area}.

With that in mind, we took  the data of Fig. 16 d, and  by
assuming $r_{eff}$ = 290 nm as in our simulation, we calculated
$\Phi \approx $3.5 eV or, with $\Phi = $5 eV, $r_{eff} \approx$
200 nm. Similar calculations made on other data samples produced
results ranging from 2 to 10 eV.

It should be noticed also that $\Phi$ can vary across the film and
can be affected by the presence of surface adsorbates or defects.

\subsection{Turn-on field}

To meaningfully compare field emitters, one usually reports the
turn-on and the threshold fields, which are respectively defined
as the macroscopic fields needed to extract current densities
respectively of  10$\mu$A$\cdot$cm$^{-2}$ and 10mA$\cdot$cm$^{-2}$
(which are the current densities required to light or saturate a
pixel in a display).

The lack of a precise evaluation of the current density forced us
to adopt a different definition of the turn-on field: following
reference \cite{turn-on}, we defined the turn-on voltage as the one
corresponding to the upward bending of the curve in the FN plot,
i.e to the establishment of the FN emission regime, and then we
estimated the macroscopic turn-on field from it.

\begin{figure}
\includegraphics[width=8.0cm]{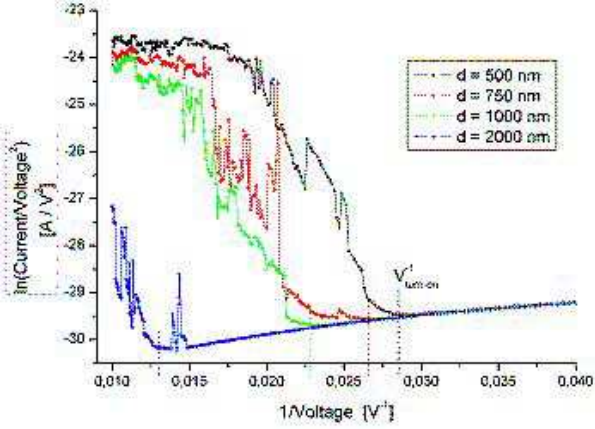}
\caption{\label{fig:epsart} Evaluation of the turn-on potential
from FN plots with different nominal tip-CNT distances. The
turn-on potential correspond to the point of upward bending. P =
10$^{-3}$ mbar, room temperature.}
\end{figure}

\begin{figure}
\includegraphics[width=8.0cm]{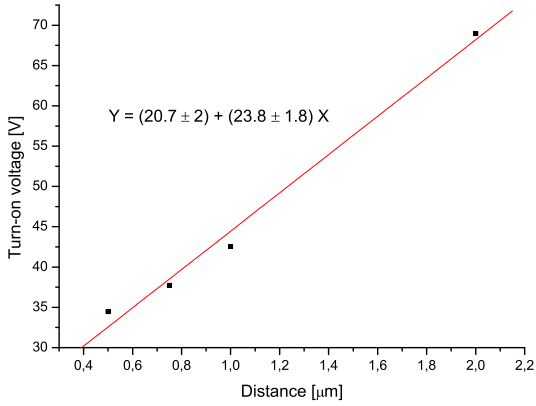}
\caption{\label{fig:epsart} Turn-on voltage as function of tip-CNT
distance.}
\end{figure}

Referring to Fig. 18, we estimated $V^{-1}_{turn-on}$ and then we
calculated E$_{turn-on}$ = V$_{turn-on}/(k_{eff}d)$ . During the
measurements, we started at $d = 1 \mu$m and then we drove the
sample by fixed steps to given distances. To rely only on those
movements, we evaluated the turn-on field with a differential
method, as $E_{turn-on}=\frac{1}{k_{eff}}\frac{\partial
V_{turn-on}}{\partial d}$ (E$_{turn-on}$ is assumed constant on
the range of $d$). Since, the distance $d$ is an unreliable parameter, we avoided it
in our quantitative analysis by relaying on the steps of the
piezoelectric displacing the sample.

A plot of V$_{turn-on}$ as a function of $d$ (Fig. 19) can help
evaluate the derivative, that equals the slope of the fitting
straight line. This algorithm provides $E_{turn-on}\approx 15
V/\mu m$, a good figure considering the screening effect, which
agrees with values obtained on similar MWCNT films (see for
example Ref. \cite{turn}).

\subsection{CNT capture}

The bias voltage induces a considerable mechanical stress on an
emitter of nanometric section. Even an applied field of few
V/$\mu$m has been proven to be sufficient to deflect and
straighten carbon nanotubes \cite{saturation1}.

The force applied during a sweep can be sufficient to reorient or
even to peel nanotubes off. Sometimes a metal layer, Ta for
example, is deposited under the catalyst to assure a better
mechanical and electrical anchorage of the nanotubes to the
substrate, a trick that we did not use.

\begin{figure}
\includegraphics[width=7.0cm]{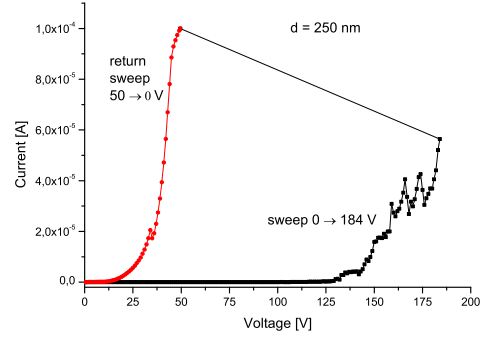}
\caption{\label{fig:epsart} I-V characteristic with the attachment
of one or more nanotubes to the AFM tip, occurring at 184 V and
producing a short circuit between the tip and the CNT film. The
red curve is the one measured with AFM probe in contact with the
CNT film. SMU current  compliance 10$^{-4}$ A.}
\end{figure}

The long nanotubes, attracted to the AFM tip, still attached to
the substrate or pulled up but still in electric contact with film
(because partially immersed in it), can create a conducting path
between the probe and the film. Such phenomenon has been
systematically observed for $d\leq$350 nm while approaching higher
voltages. An example is shown in Fig. 20: at 184 V, when a FE
current of 57 $\mu$A is flowing, one or more CNTs are peeled off
and attached to the AFM tip, causing a low resistive path between
the probe and the film underneath, provoking a current jump to its
compliance value (10$^{-4}$A). During the return sweep, below 50
V, the current follows the typical path it has with the probe in
electric contact with the CNT film surface.

The tensile force applied on an infinitesimal area dA by the electric
field E$_S$ can be expressed as $dT=(\varepsilon _0/2)E^2_SdA$.
With the parameters of the experiment in Fig. 20, $E_S \approx
\gamma \frac{V}{d\cdot k}\approx 1.4\cdot 10^4 \frac{V}{\mu m}$ ,
a stress of $\sim 8.7 \cdot 10^{-4} N/\mu m^2$is applied. Hence,
in correspondence of the emitter failure voltage, the tensile
force on a typical nanotube of 30 nm diameter is  $\sim 2.5 \mu$N.

\subsection{Time  stability}

The study of the emission current over time is very important for
the utilization of CNTs in technological applications
\cite{stability}.

We measured the stability of field emission over
periods of several hours, both in the high and low current regime.

Individual nanotubes are known to have high instabilities compared
to dense films were current is averaged on an extremely large
number of emitters \cite{film}. In our setup, current is
contributed by a limited number of emitters and for lack of
"ensemble effect" enhanced fluctuations are expected.

\begin{figure}
\includegraphics[width=7.0cm]{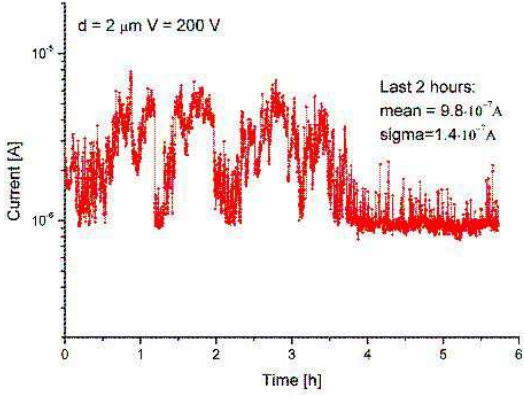}
\caption{\label{fig:epsart} Field emission current vs time.
Applied Voltage 200V, tip-cnt distance 2$\mu$m. P = 10$^-7$ mbar, room temperature.}
\end{figure}
\begin{figure}
\includegraphics[width=7.0cm]{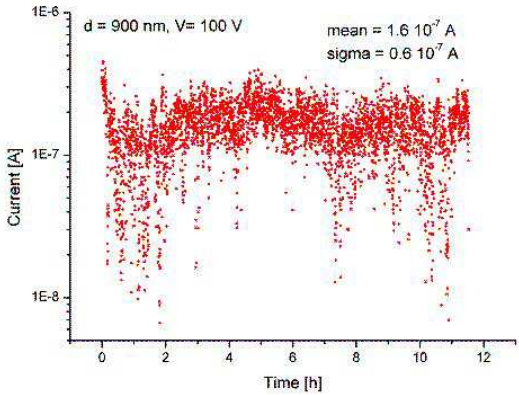}
\caption{\label{fig:epsart} Field emission current vs time.
Applied Voltage 100V, tip-CNT distance 900 nm. P = 10$^-7$ mbar, room temperature.}
\end{figure}

Fig. 21 shows the result obtained for emitters in the high current
(or saturation) regime. A measurement was taken every 5 seconds.
During the first 4 hours the current was fluctuating over an order
of magnitude, with high frequency variations superposed to a sort
of low frequency oscillation; a good stability of 15 \% (one
sigma), at the lowest current, was achieved afterwards.

\begin{figure}
\includegraphics[width=8.0cm]{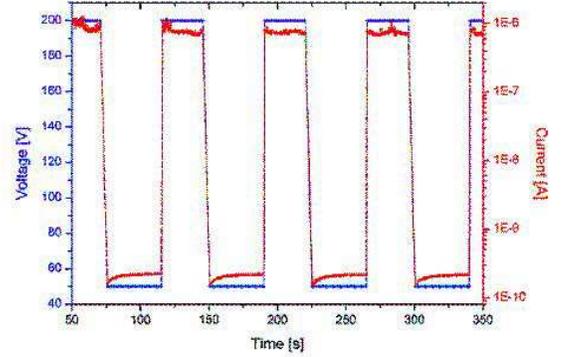}

\includegraphics[width=8.0cm]{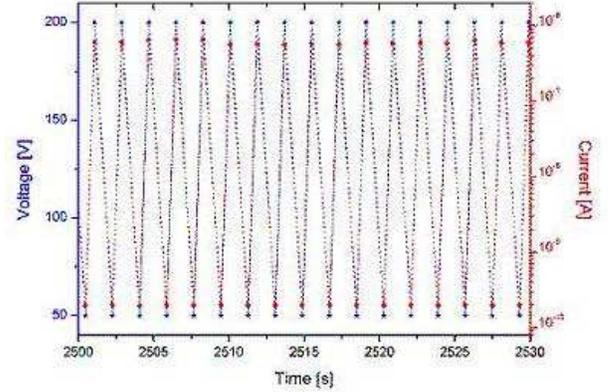}
\caption{\label{fig:epsart} Field emission current upon
application of a pulsed and switching voltage.P = 10$^-7$ mbar, room temperature.}
\end{figure}

It should be pointed out that operating the nanotubes in the
current saturation regime is always risky as the current
saturation is a sign of power dissipation and therefore of
possible degradation, that might have been occurred during our
measurement before stabilization.

In low current (non-saturation) regime, no overall decay in the emission
current was observed over a
period of 12 hours, as shown in Fig. 22 (where the current was
measured every 20 s for 12 consecutive hours). A one sigma
stability less than 30 \% was achieved.  This can be considered a
good result compared to reported instabilities up to 50 \% in the
same regime \cite{instability1,instability2}.

To investigate stability of field emission upon a changing voltage
we measured the current with a low frequency pulsed and switching
voltage for a few hours, obtaining results as those shown in Fig.
23. The FE current was found to follow the variations of the
applied voltage over a time of about 3 hours without any failure.

\subsection{Field emission under laser irradiation}

The effects of laser irradiation on FE from MWCNTs is very
interesting for the study of their optoelectronic properties and
for their possible utilization in radiation detectors
\cite{ir1,ir2}. Reference \cite{laser}, for example, reports a
marked increase of emission current with the irradiation duration
(more that a factor 15  after 6 min irradiation with continuous wave (CW) 633 nm
laser focused to a 5 mm spot size and with fluence of 10 mW) and
suggests a laser induction of surface plasmons as possible
explanation.
\begin{figure}[h]
\includegraphics[width=7.0cm]{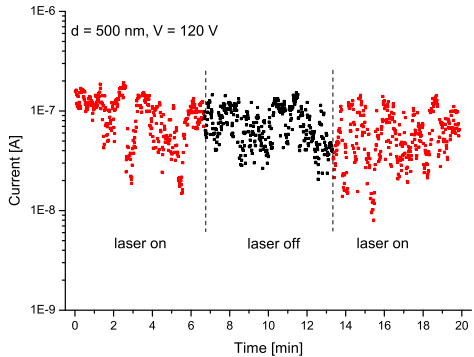}
\caption{\label{fig:epsart} FE current measured at given distance
and voltage with and without laser irradiation of the emitting
surface.}
\end{figure}
\begin{figure}[h]
\includegraphics[width=7.0cm]{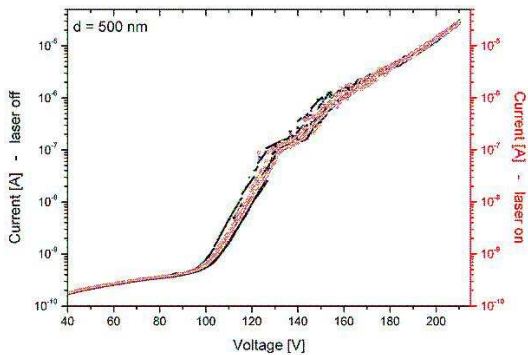}
\caption{\label{fig:epsart} I-V characteristic with and without
laser irradiation of the emitting surface. Several voltage sweeps
showed no significant FE current difference.}
\end{figure}
To investigate laser effects, we compared FE current with and
without irradiation on several positions, after electric
conditioning. CW lasers with wavelengths of  655 nm and 670 nm, 5
mW power and spot size of about 5 mm were used (the fluence was
well below the threshold for destructing the CNTs). Comparison of
signals with and without lightening (laser beam was almost
parallel to the sample surface) did not show any significant
difference, neither as a function of the irradiation duration
neither as a function of the bias voltage (Fig. 24 and 25).
However, we consider this experiment non-conclusive and we plan to
repeat it with more powerful and different wavelengths lasers.

\section{Summary}

We have reported measurements of field emission from a vertical
and quasi-aligned CNT film, produced by catalytic CVD, by using an
original measurement setup based on a voltage biased AFM/STM
nanometric  probe. We were able to accurately characterize the
local field emission behaviour, with currents averaged on a
limited number of emitters. We have found that a modified Fowler
-Nordheim model accounting for a series resistance provides a
satisfactory explanation of the experimental results. We also gave
an estimation of relevant parameters as field enhancement factor,
MWCNT workfunction and turn-on field. We studied FE stability and
effects of red laser light.

\begin{acknowledgments}
We thank R. Fittipaldi and A. Vecchione for the SEM images and
GINT (Gruppo INFN per le NanoTecnologie) collaboration for
motivating and supporting this work.
\end{acknowledgments}


\begin{thebibliography}{99}
\bibitem{MAX} MAXWELL is a software by Ansoft Corporation (Pittsburg,Four
Station Square, Suite 200, PA 15219) which solves two or three
dimensional electromagnetic problems through finite element
analysis
\end{thebibliography}
\end{document}